\documentclass[review]{elsarticle}

\usepackage{hyperref,graphicx,amssymb,float,amsmath,color,latexsym,appendix}

\begin{document}

\begin{frontmatter}

\title{\large{The illiquidity network of stocks in China's market crash}}




\author[label1]{Xiaoling Tan}
\author[label1]{Jichang Zhao}

\address[label1]{School of Economics and Management, Beihang University, Beijing, China}
\cortext[cor_label]{Corresponding author: jichang@buaa.edu.cn}

\begin{abstract}
The Chinese stock market experienced an abrupt crash in 2015, and over one-third of its market value evaporated. Given its associations with fear and the fine resolution with respect to frequency, the illiquidity of stocks may offer a promising perspective for understanding and even signaling a market crash. In this study, by connecting stocks with illiquidity comovements, an illiquidity network is established to model the market. Compared to noncrash days, on crash days, the market is more densely connected due to heavier but more homogeneous illiquidity dependencies that facilitate abrupt collapses. Critical stocks in the illiquidity network, particularly those in the finance sector, are targeted for inspection because of their crucial roles in accumulating and passing on illiquidity losses. The cascading failures of stocks in market crashes are profiled as disseminating from small degrees to high degrees that are usually located in the core of the illiquidity network and then back to the periphery. By counting the days with random failures in the previous five days, an early signal is implemented to successfully predict more than half of the crash days, especially consecutive days in the early phase. Additional evidence from both the Granger causality network and the random network further testifies to the robustness of the signal. Our results could help market practitioners such as regulators detect and prevent the risk of crashes in advance.
\end{abstract}


\end{frontmatter}


\section{Introduction}
\label{sec:intro}

The stock market plays the most profound role in the financial systems of modern economies such as China. An abrupt stock market crash, such that in 2015 in which approximately 15 trillion yuan in wealth evaporated, could therefore be a catastrophic shock to economies and bring about enormous losses to society as a whole. In fact, how to understand market crashes and implement early warnings has been an important issue and trending topic not only in finance but also in interdisciplinary fields since the crisis. However, it is conventionally understood that a market crash might be a typical black-swan event, which is nearly impossible to predict due to sophisticated factors beyond the crash and unexpected entanglements with external systems. Nevertheless, the associations between investor behaviors, such as expectations, emotions and imitations, and market performance, especially their power in return predictions \cite{ZhouTales,wang2019aggr}, imply that trading behaviors may provide a new but promising perspective in probing and generating warnings of a market crash. It is accordingly reasonable to assume that in the stock market, the behaviors of investors, in particular irrational behavior, would help disseminate disturbance, amplify panic and then result in a crash. In particular, the details of every trading decision in high-frequency records further offer a big-data proxy to investigate the collective behavior of investors, either before, during or after a market crash.

Liquidity, referring to the spread between the bid price and ask price, inherently reflects the expectations of investors regarding the future performance of stocks in their elementary trading decisions. Illiquidity, which inversely originates from the pessimism of investors, would thus increase crash risk since it dissolves effective price information and disseminates panic throughout the market. Given the significant impact of investor emotions, especially negative emotions \cite{ChiuInvestor,FloriCommunities}, illiquidity can also be contagious, e.g., investors frightened by stock illiquidity tend to sell the other stocks they hold to maintain their own liquidity, reluctantly generating more illiquid stocks. Intuitively, the above contagion thus results in the comovement of illiquidity and offers paths for the cascading of illiquidity throughout the market. Hence, to model a market crash from a systemic perspective, it would be natural to connect stocks with similar illiquidity fluctuations and build a network to represent the market. Therefore, in the established illiquidity network, links among stocks represent the possibility of cascading crashes throughout the market, suggesting a new perspective for profiling market crash dynamics. Although it is not a new idea to transform a market into a network, linking stocks in terms of illiquidity is rarely done. More important, different from previous networking models of mutual fund sharing \cite{LuHerding}, price comovements~\cite{MantegnaHierarchical} or the Granger test~\cite{wang2018corr}, illiquidity can be captured dynamically at a fine resolution, i.e., at the minimum decision granularity of bid and ask prices and inherently associated with investor emotions such as panic. This means that in terms of elementary decisions in trading and their contagion, the illiquidity network provides a microperspective on market crashes.

Although there are many studies on stock market crashes, the results on crash forecasts remain inadequate, and more effort is desperately needed. Unlike many emerging financial markets, however, the Chinese stock market is unusual since it is dominated by individual investors, who execute most of the daily trading~\cite{ShenzhenNews}. Contrary to their institutional counterparts, individual investors are more emotional and susceptible, meaning they are more likely to be frightened, spread panic and overreact to external disturbances. They even imitate trading strategies and help create herding in the market. These characteristics might undermine the challenges that make crashes difficult to predict and suggest the possibility of detecting crashes in the Chinese market at an early stage. In terms of illiquidity, trading behaviors in extreme market situations can be finely examined from a microperspective, helping identify the sources of market volatility and extreme stock price movements. In addition, anomalies in the evolution of illiquidity networks can also be probed based on the differences between crash days and noncrash days, which paves the way for developing warning signals of market crashes.

Inspired by the above intuitions and motivations, the aim of this study is to profile and generate a warning system for Chinese market crashes using an illiquidity network. The illiquidity of stocks is defined and derived from 2.3 billion trades in 2015, from which profound associations between illiquidity and negative emotions of investors such as fear are disclosed. The illiquidity dependency between stocks, measured by mutual information, can surprisingly distinguish crash days from noncrash days. It is also interesting that the market is more connected and homogeneous due to heavier and less-deviated illiquidity dependencies on crash days. In the illiquidity network, influential stocks in crashes are found to be those with large capital values or belonging to the finance sector. The dynamics of the crash are also profiled in the illiquidity network as cascading failures of losing illiquidity from stocks with smaller degrees to those with higher degrees that are usually located in the core and then spread out to the fringe. More important, an early signal, which simply counts the days without systemic failures within a window of the previous five days, is presented to accurately warn of more than half of the crash days in 2015. Further evidence from networks constructed either from Granger causality between illiquidity series or randomized ask-bid sequences testifies to the robustness of the signal. Our results decently demonstrate the power of the illiquidity network in understanding Chinese market crashes and could help practitioners, particularly regulators, inspect risky stocks and prevent possible crashes in advance.

The remainder of the paper is organized as follows. Section 2 reviews the related literature. Section 3 introduces our datasets and the methodology for measuring illiquidity. Section 4 presents the results from illiquidity networks, and Section 5 tests the robustness of the results. Section 6 concludes the paper with a brief summary and highlights the limitations of our study that provide promising directions for future research.

\section{Literature review}
\label{sec:lr}

Due to the late development of China's stock market and the obvious gap with developed foreign markets, there some unique features of the Chinese stock market have been discussed among academic scholars and practitioners. On the one hand, Yao et al. indicated that Chinese investors exhibit different levels of herding behavior \cite{YaoInvestor}. On the other hand, Xing and Yang found that increased correlation among stocks could ignite market crashes \cite{XingHow}. Further, Tian et al. found that institutional investors (primarily pension funds) provide a stabilizing effect during extreme market down days \cite{TianWho}, unlike Dennis and Strickland, who revealed that institutional investors magnify extreme market movements by buying (selling) more on return up (return down) days in U.S. markets \cite{DennisWho}. Although there are many related studies of either the Chinese market or foreign markets, no detailed explanations or early warning signals of stock market crashes have been provided to mitigate risks. Moreover, the dominance of individual investors in the Chinese market also implies possible abnormalities in trading behaviors that can be sensed and detected as warnings before a crash. This progress inspires and motivates us to depict crashes in China's market through perspectives that are inherently associated with investor behavior, particularly trading decisions.

In fact, previous efforts have already suggested that illiquidity is closely related to stock market crashes. Amihud et al. presented evidence linking the decline in stock prices to increased illiquidity by examining the bid-ask spread during market crashes \cite{Amihud1990Liquidity}. As the return is more comparable to price, related research on associations between return and illiquidity has rapidly grown. Amihud and Bekaert et al. stated that there is a positive correlation between stock returns and liquidity in terms of the daily ratio of absolute stock return to its dollar volume and the proportion of zero return days \cite{Amihud2002Illiquidity,BekaertLiquidity}. 
Furthermore, Nagel indicated that the main reason for the evaporation of liquidity during crashes is the increasing expected returns of liquidity \cite{NagelEvaporating}. Moreover, measuring illiquidity, e.g., through the bid-ask spread, is deeply rooted in the minimum decision granularity of daily trading and thus can be inherently derived from high-frequency trading records of investors. Additionally, illiquidity contains future economic information that can be employed for stock market forecasting \cite{StollInferring,ChenMicro}. Therefore, it is feasible to explore stock market crashes from the perspective of illiquidity, but existing studies still lack explanations, cascading dynamics, and warning signals of a crash.

Illiquidity may also be influenced by both internal and external factors, including stock attributes, policies and industry, which should be considered when attempting to understand market crashes. Stoll et al. suggested that stock attributes such as market value, volume and volatility can significantly reshape stock illiquidity \cite{StollInferring,ChenMicro,ChordiaMarket}. On the other hand, An et al. found that macroeconomic factors such as media independence, policy uncertainty, default risk and funding conditions have a remarkable impact on illiquidity \cite{AnTheImpact,ChungUncertainty,BrogaardStock,BrunnermeierMarket}. This evidence implies that stocks can be well profiled in terms of illiquidity and, more important, that external shocks to the market can also be absorbed and thus sensed through illiquidity. In addition, the illiquidity of individual stocks covaries \cite{ChordiaCommonality,HasbrouckCommon,HubermanSystematic,AcharyaAsset,DengForeign}, suggesting, in essence, that illiquidity can be contagious in the market.

Modeling the market as a network of stocks to examine crashes is a new and promising approach in recent efforts. Traditional market networks were mainly forged through correlations among stocks. For example, Mantegna linked stocks through linear correlations between price series~\cite{MantegnaHierarchical}, Kenett et al. and Wang et al. defined connections through partial correlations that better capture the nonlinear associations among stocks~\cite{KenettDominating,wang2018corr}, and Wange et al. employed the Granger test to introduce risk causality into the networks~\cite{wang2017extreme}. Stocks can also be connected due to common investors~\cite{LuHerding}. By removing failed stocks, e.g., reaching the down limit and suspending transactions, a market crash can then be reflected through the collapse of the network. The evolution of the network topology before and after the 2008 financial crisis in the South African, Korean and Chinese stock markets was investigated in past work \cite{MajapaTopology,NobiEffects,YangAnalysis}, in which minimum spanning trees (MSTs) were carefully examined. Li and Pi proposed a complex network-based method to understand the effects of the 2008 global financial crisis on the main global stock index \cite{LiAnalysis}. In addition, Bosma et al. used network centrality to identify the position of the financial industry in the network, which can be a significant predictor of bailouts \cite{BosmaToo}. In particular, the turbulence in the years 2015--2016 was investigated by transforming the Chinese stock market into a complex network, and the results showed that influential stocks and sectors existed within the market crash \cite{LuHerding,khoojineNetwork}. Nevertheless, connecting stocks because of illiquidity associations is rarely considered when constructing the market network. The failure of existing studies on market crashes to establish illiquidity networks allows for the development of new perspectives in this paper.

In summary, although extensive efforts have been devoted to the association between stock illiquidity and market crashes, few insights are available on illiquidity networks based on high-frequency transaction data. Given the closeness between stock illiquidity and both the internal and the external factors of the market, examining crashes from the perspective of illiquidity networks could offer more insightful observations and inspiration. Moreover, the dominance of individual investors in the Chinese stock market also indicates that a trading abnormality, which can be captured by illiquidity and its contagion at a fine resolution, could produce novel warning signals of risks before a crash. From an interdisciplinary perspective, a big-data proxy based on highly extensive trading records from before, during and after the 2015 crash of China's stock market will be employed to measure illiquidity, establish networks, examine crash dynamics and detect warning signals.

\section{Dataset and methods}
\label{sec:data}

\subsection{Dataset}
\label{subsec:data}

The data sample employed in this study consists of stocks selected from the Shenzhen Stock Exchange and the Shanghai Stock Exchange in 2015, i.e., more than 2,500 stocks and a total of 244 trading days. In particular, transaction records of the minimum trading decision granularity include the ask price, ask volume, bid price, and bid volume for every second of every stock. The dataset is provided by Wind Information (Wind Info), a leading integrated service provider of financial data in China.

\begin{figure}[htp]
		\centering
		\includegraphics[width=120mm]{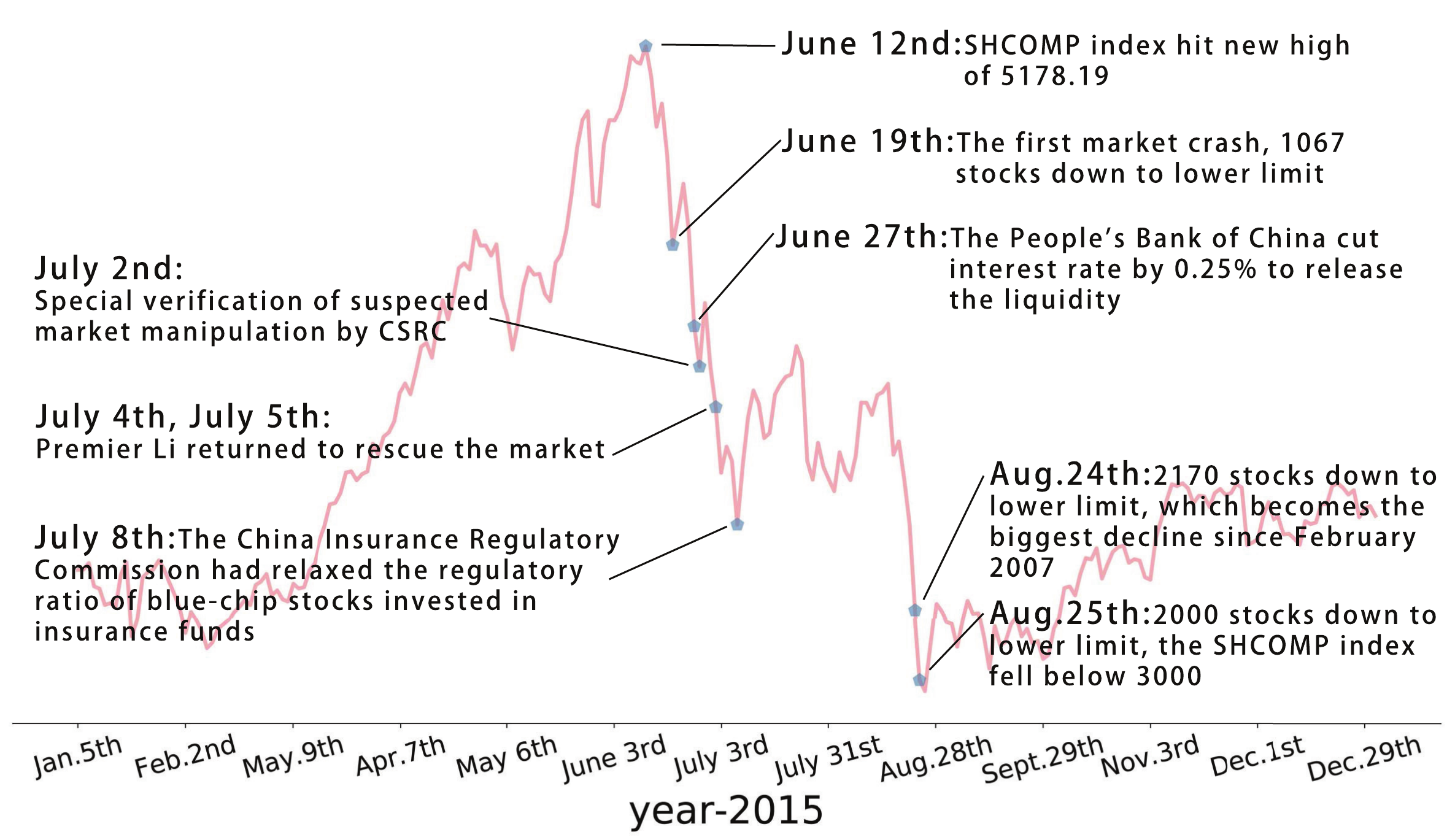}
		\caption{Review of key events of the market crash in 2015.}
		\label{fig:crash_event}
\end{figure}

To identify a stock market crash, crash days are defined as those days when the number of stocks being sold to the down limit (the allowed maximum one-day drop of a stock, i.e., ten percent of its closing price on the last day) exceeds than a specific threshold. The determination of the extent to which the market could be defined as crashed depends on the tradeoff between the number of crash observations (days on which the market crashes due to systemic failure) and the likelihood of detecting early warning signals. It is intuitive that a lower threshold would result in more crash days, while more noisy crashes caused by random failure might be included. In contrast, a higher threshold would lead to fewer crash observations, with most being caused by systemic failure, while having too few samples cannot guarantee the accuracy and significance of warning signal detection. Hence, the object in setting the threshold is to select an appropriate number of observations that have the most crashes caused by systemic failure and, accordingly, could help fairly and convincingly judge the approach of early signal detection. As shown in Appendix Fig.~\ref{number_of_crash_days}, the number of crash days gradually decreases as the threshold increases, which fits our intuition. Note that the decrease in crash observations shows an abrupt change as the threshold approaches 800, particularly when comparing the nearly equal days for 600 and 700 days, suggesting that 800 days may be an inflection point. More important, our later experiments demonstrate that the early signal cannot be established when the threshold is less than 600 (or 700) due to an excessive number of noisy observations, and the prediction accuracy of detection also deteriorates because of too few observations when the threshold is greater than 800 (e.g., 900).

Consequently, we chose 800 as the threshold in this study. Specifically, as seen in Fig.~\ref{fig:crash_event}, in 2015, there were 17 trading days on which the stock market crashed, including June 19th, June 26th, June 29th, July 1st, July 2nd, July 3rd, July 6th, July 7th, July 8th, July 15th, July 27th, August 18th, August 24th, August 25th, September 1st, September 14th, and October 21st. Other days before or after these crash days will be defined as noncrash days and consist of the counterparts for further comparison. Note that the threshold selection might be market-dependent, but it barely impacts the framework itself later presented in this study to detect crash warnings at early stages. In line with this, the framework is robust to the threshold setting and could be expanded to other markets.

\subsection{Measuring illiquidity}
\label{subsec_mi}

The transaction data are full of noise due to the frequent occurrences of quotes. To filter out noise and smooth the data, a fixed time window has to be selected to average the spread. In addition, the window length is determined by the tradeoff between data frequency and computational costs. Different from previous methods of constructing stock networks, the building of illiquidity networks is based on high-frequency records to reflect the minimum decision unit in investment. It should be implemented to capture the microcosmic investment footprints from high-frequency information. However, the original frequency of our raw data is recorded in seconds, and the occurrence time of each transaction is inconsistent, leading to the necessity of unifying the time measurement to align the data length after calculating the spread. To align the length of spread sequences, time windows of seconds (e.g., 15 s, 30 s, 60 s, 90 s) have been set to merge the raw data, and obviously shorter windows can produce higher-frequency and better-resolved granularity in averaging. For high-frequency data, shorter windows also entail substantial calculations. Because of the above considerations, we set the time window as one minute, which is a better option to maintain the high-frequency information contained in the raw records as much as possible and simultaneously make the calculation of the liquidity indicator feasible. In addition, time windows of more than one minute also lead to too short spread sequences and possibly result in biased mutual information due to the insufficiently sampled distribution of illiquidity during the inference of illiquidity networks in later experiments. However, note that compared to the previous study, one minute is still short enough to reflect the investment behavior of investors at the smallest decision granularity. In addition, it is necessary to convert the length of the data sequence into 237 minutes for every stock in a day because the Shenzhen Stock Exchange has adopted a collective bid for the last three minutes.

Various methods have been presented to calculate illiquidity for different occasions and purposes. The methods applied to low-frequency data work well when high-frequency data are not available \cite{Amihud2002Illiquidity,LesmondA,RollA,CorwinA,ChengLiquidity}; nevertheless, it is clear that approaches based on high-frequency data perform better due to richer information and greater accuracy \cite{ChungUncertainty,GoyenkoDo}. Here, illiquidity is expected to capture the minimum decisions in trading behavior; hence, the bid-ask spread based on high-frequency records, which is always considered the best method, is selected to measure illiquidity \cite{EasleyChapter,KyleContinuous}. Moreover, it is known that the size of the transaction has a substantial impact on illiquidity, and we further update the measure by adding the quoted amount as the weight of the spread. Illiquidity can be denoted as

\begin{equation}
\label{eq:ill}
I_t=({\sum_{i=1}^{10} A_{it}}V_{it}/{\sum_{i=1}^{10} V_{it}}-{\sum_{j=1}^{10} B_{jt}V_{jt}}/{\sum_{j=1}^{10} V_{jt}})/P_{mid,t},
\end{equation}
where $A_{it}$ is the ask price of investor $i$ at time t, $V_{it}$ is the ask volume of investor $i$ at time $t$, $B_{jt}$ is the bid price of investor $j$ at time $t$, $V_{jt}$ is the ask volume of investor $j$ at time $t$, and $P_{mid,t}$ is the mean ask price and bid price at time $t$. The value of 10 is derived from the raw data, because in total 10 ask prices and bid prices are separately offered in the data records. The definition indicates that the lower the weighted spread is, the lower the transaction cost and the lower the illiquidity.

\begin{figure}[htp]
		\centering
		\includegraphics[width=120mm]{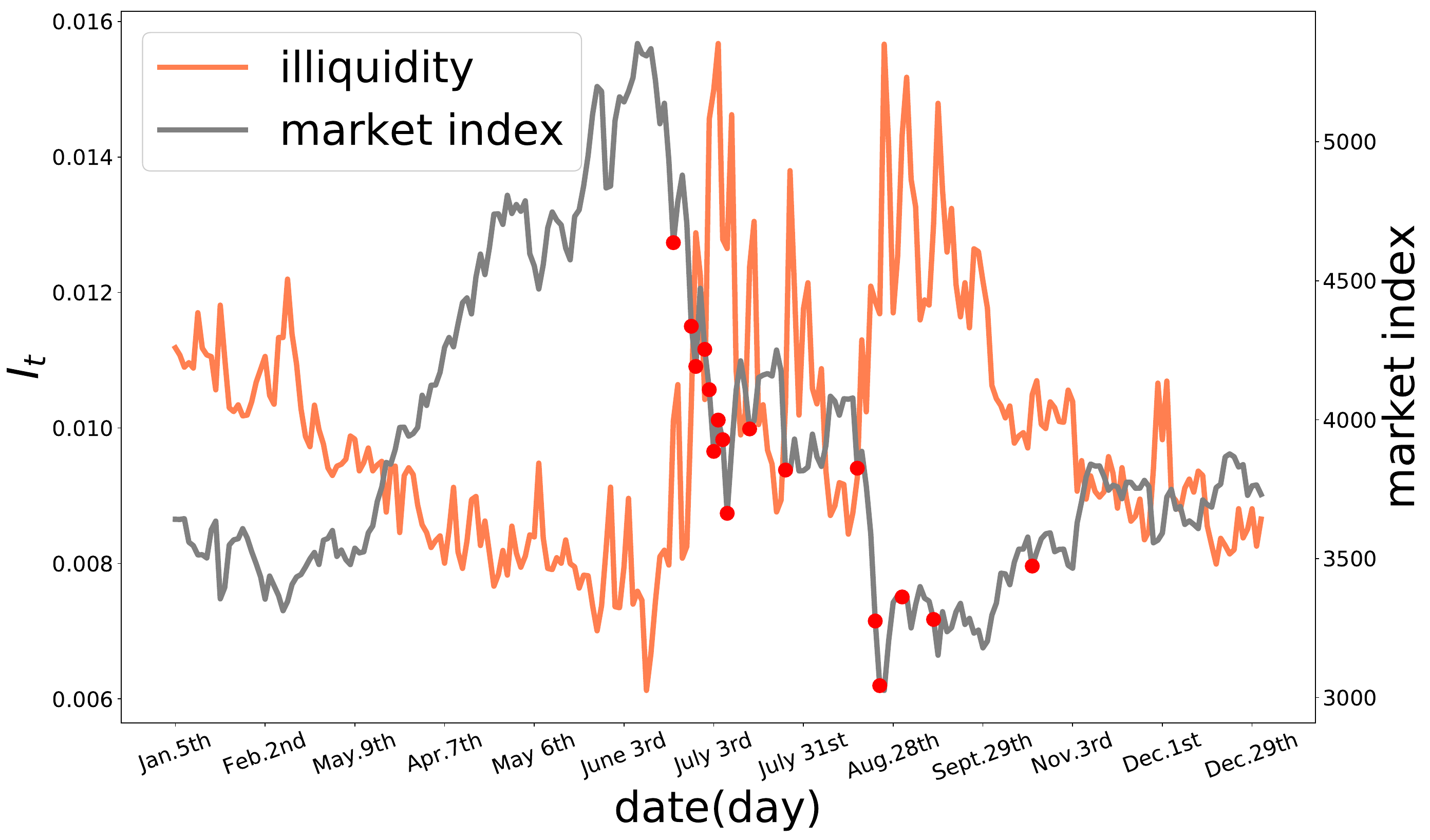}
		\caption{Illiquidity with stock index. $I_t$ is the illiquidity we measured, and the market index represents the CSI 300 Index. The correlation between illiquidity and the market index is -0.64 with a $p$-value of 0.00. The red dots indicate the crash days.}
		\label{fig:iindex}
\end{figure}

The potential capability of illiquidity in understanding market crashes is simply illustrated in Fig.~\ref{fig:iindex}, in which the market index is negatively associated with the fluctuation of illiquidity we measured. China's stock market experienced a period of ups and downs in 2015, in which more than ten crash days erupted in succession. Specifically, illiquidity continued to decrease before June, and at this stage, investors easily completed transactions due to reduced costs, and the market index continued to soar. In contrast, illiquidity demonstrated an abrupt increase in June and August, when crash days are common and resulted in high transaction costs, inactive investors and a falling market index. 
These observations confirm the previously disclosed association between illiquidity and crashes in China's stock market and inspire the following investigations from the novel perspective of illiquidity networks.

\section{Results}
\label{sec:results}

\subsection{Illiquidity and crashes}
\label{subsec:illcrash}

\begin{figure}[htp]
		\centering
		\includegraphics[width=120mm]{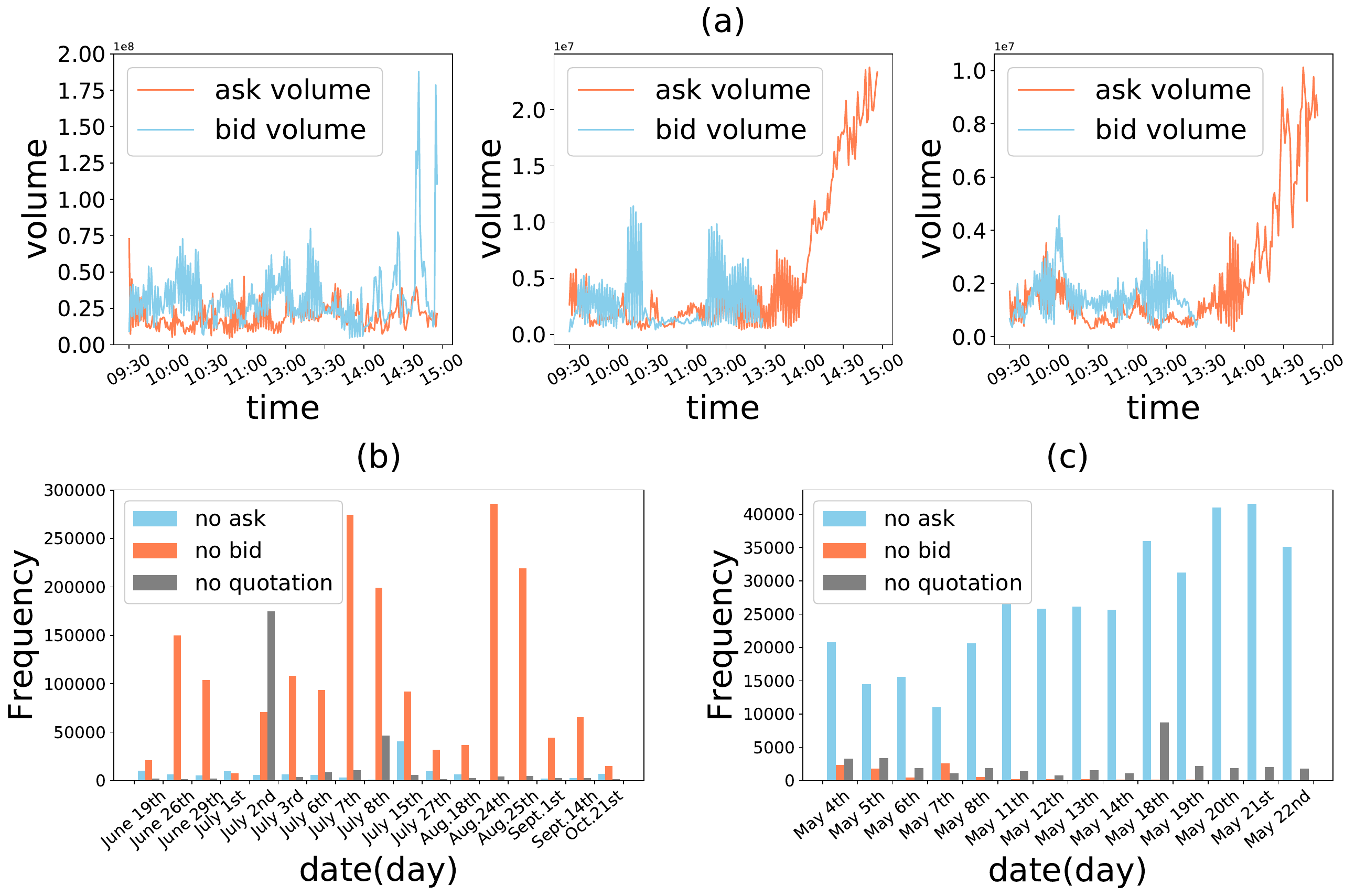}
		\caption{Trading behaviors on crash and noncrash days. (a) shows the ask and bid volume on the market crash day of June 26th; the stocks are randomly selected from the sample. The first subgraph shows the stocks not losing liquidity on the crash day, and the other two show stocks losing liquidity on the crash day. When one of the asks or bids does not exist, or neither of them exists, then the stock loses liquidity. (b) shows the quotations of buyers and sellers on crash days, where the frequency is defined as how often each action occurred in every minute of the crash day, no ask means no buyers quote and no bid means no buyers quote. (c) shows the quotations of buyers and sellers on noncrash days.}
		\label{fig:tf_crash}
\end{figure}

It is assumed that trading behaviors, especially elementary actions such as asks and bids in high-frequency data, would be fundamentally influenced by shocks such as market crashes. As Fig.~\ref{fig:tf_crash}(a) shows, when stocks approached the down limit on crash days, the volume of bids experienced an abrupt decline and then vanished; in contrast, the ask volume soared, implying that many investors were forced to sell shares owing to panic selling and risk prevention. Note that, here, these randomly selected stocks are only employed to produce an intuitive and vivid picture of how liquidity varies from noncrash to crash days in the raw data. However, approaching the down limit might also occur on noncrash days. To further test the impact of market crashes on trading behaviors, we randomly select ten crash days and noncrash days to compose two different groups and compare the occurrence frequencies of no-ask, no-bid, and no-quote when stocks experienced a lower limit. In an unexpected and surprising finding, crash days can be distinguished from noncrash days. 
Specifically, as seen in Fig.~\ref{fig:tf_crash}(b), no quotations, which would result in liquidity loss, mainly come from no-bids on crash days instead of no-asks on noncrash days. This opaque impact of market crashes on trading behaviors further suggests that illiquidity, the calculation of which is based on both asks and bids, would inherently capture the potential for market crashes from the novel perspective of trading decisions.

\begin{figure}[htp]
		\centering
		\includegraphics[width=120mm]{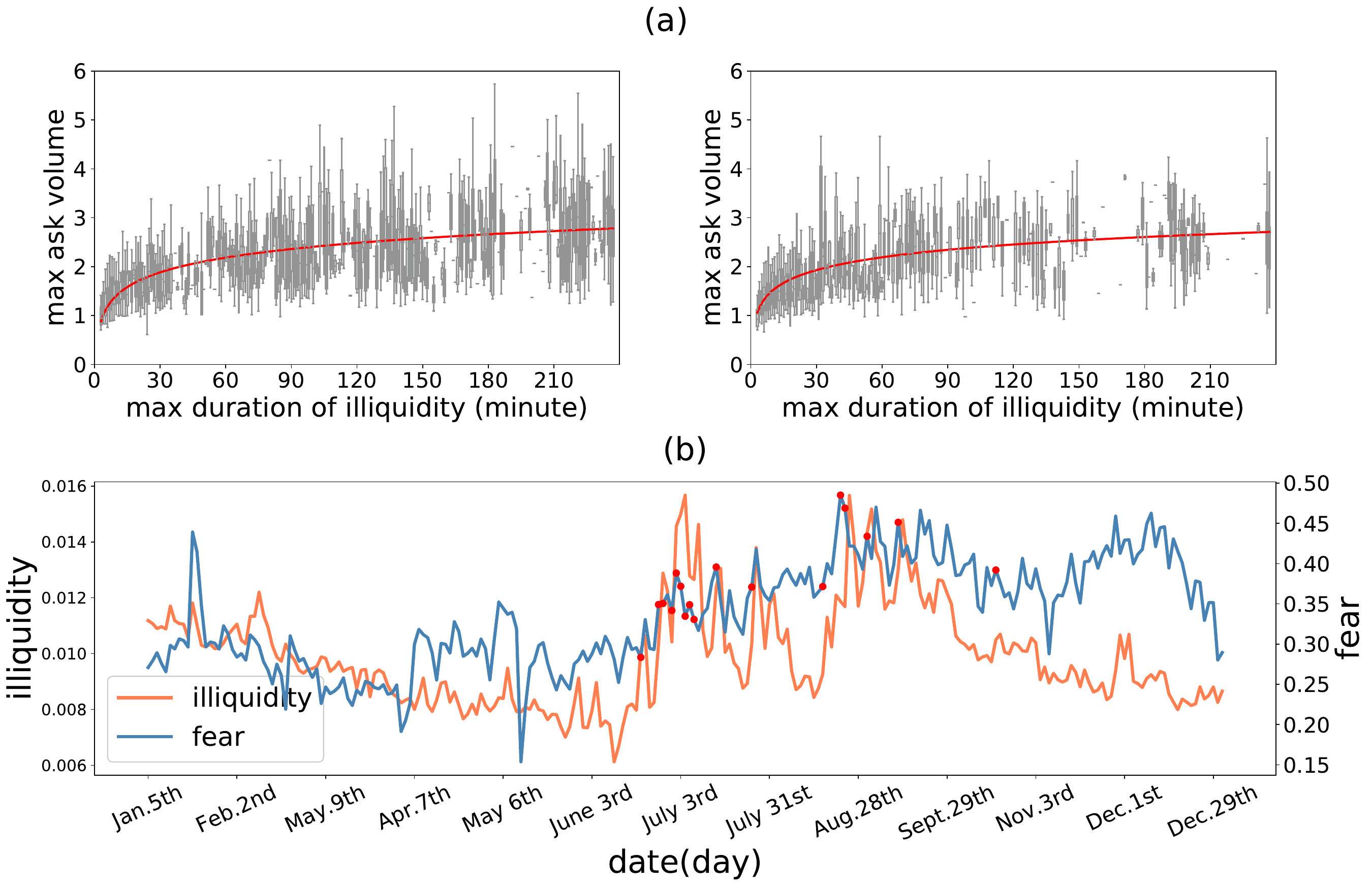}
		\caption{Maximum duration of illiquidity due to no bid. (a) shows the correlation between the maximum ask volume and the maximum duration of illiquidity, which indicates the longest duration for which liquidity is lost. Note that there may be several periods of liquidity loss per stock in a day. The stock is randomly selected from the sample, and the other stocks have similar relationships. (b) shows the correlation between investors' fear and illiquidity, whose value is 0.44 with a $p$-value of 0.00.}
		\label{fig:max_duration}
\end{figure}

A zero volume of bids but a soaring number of asks suggests that on crash days, investors are anxious, and their anxiety accumulates. As Fig.~\ref{fig:max_duration}(a) interestingly shows, the maximum volume of asks grows logarithmically with the duration of illiquidity loss, i.e., no-bid. In fact, Subrahmanyam proposed a theoretical model suggesting a magnet effect whereby circuit breakers may increase price variability and exacerbate price movements when the price is very close to the trigger level, potentially leading to increased trading volume~\cite{Subrahmanyam}. This model explains the increase in trading volume before the stock price approaches the trigger. On the other hand, Lee et al. examined the duration of the post-halt effect and found that volume and volatility increase rapidly after the halt~\cite{leevolume}, which is also consistent with our results. Moreover, our results further indicate that the maximum ask volume increases logarithmically as a function of the duration of illiquidity, which validates the existence of the magnet effect in the Chinese market and enriches its details with respect to quotes. Due to the existence of the circuit breaker, the trading volume will increase rapidly near the critical point. Nevertheless, market orders would be imbalanced since sellers are more inclined to sell assets while buyers choose to wait. At this stage, investors can be easily affected by others, especially by the spread of pessimism. It further suggests that the illiquidity rooted in high-frequency records could capture investment decisions such as panic selling. Given the possibility of sharing investors, the illiquidity of one stock that makes its investors panic due to a loss of liquidity might even make them sell other stocks they hold~\cite{lusmall}, which accordingly results in the comovement of illiquidity of different stocks and offers paths for the dissemination of illiquidity throughout the market. This logarithmic-like relationship also indicates that the longer the no-bid period lasts, the more anxious investors become and the substantial increase in the ask eventually slows. The saturation of ask volume can be explained by investors becoming less panicked when more information is obtained. From this perspective, trading actions such as an ask can also be directly connected to investor emotions, and intuitively, illiquidity based on the spread between asks and bid should be coupled with emotions, especially negative emotions.

To empirically verify the possible associations between illiquidity and investor emotions, the correlations between illiquidity and investor emotions identified on social media are examined. The data source for investor emotions comes from Zhou et al.~\cite{ZhouTales}, and these data were also employed by Wang et al.~\cite{wang2019aggr}. Specifically, investor emotions were measured through stock-related tweets on Weibo, a Twitter variant, which is also the most popular social media platform in China. A classifier was trained based on a manually labelled corpus in~\cite{ZhouTales} to automatically group each stock-related tweet (those that mentioned keywords related to the stock market) into one of five emotions, including anger, joy, sadness, disgust and fear. Those without a significant tendency in emotional expression (i.e., the odds of belonging to the above five emotions were evenly distributed) were also omitted as neutral tweets. Then, according to the posting time, tweets of a certain emotion can be aggregated to calculate the share of the corresponding emotion in all emotional tweets, which is then defined as the emotion index. One of the possible limitations of inference made based on the emotion index is sampling bias; e.g., we cannot ensure that all stock investors only express their emotions toward stocks and the market through Weibo. It is possible that they express their feelings and opinions through other outlets, such as online forums or offline conversations. However, evidence from previous efforts in which the emotion indexes were employed to competently predict stock price trends demonstrates that these emotion indexes carry realistic expectations from investors and can be convincing signals to anticipate future stock prices. Specifically, Zhou et al. revealed the Granger causality between emotion indexes~\cite{ZhouTales} and stock returns, and Wang et al. further found that these indexes indeed offer more power than other factors in stock prediction~\cite{wang2019aggr}. Hence, we argue that these efforts could ensure the reliability of these emotion indexes in representing investors' emotions toward market status.

However, it is challenging to measure and aggregate emotions toward individual stocks due to the extreme sparsity of stock-related tweets on Weibo~\cite{ZhouTales}. For the same reason, it is also impossible to calculate emotion indexes in the fine granularity of minutes or hours, meaning that emotions cannot be manifested as high-frequency indicators, unlike illiquidity. Hence, similar to previous efforts~\cite{ZhouTales,wang2019aggr}, here, we build emotion indexes for the entire market instead of a particular stock at the daily level. The averaged sequence of illiquidity in the market is accordingly aggregated into a daily sequence, and its significant associations with fear can be found in Fig.~\ref{fig:max_duration}(b). The cointegration regression also  supports its accuracy since the coefficient of determination is greater than 0.7. The positive correlation implies that illiquidity can well reflect, even at a better resolution, the fear in China's stock market, where individual investors dominate trading. It is difficult for individual investors to be completely rational; they usually blindly follow other actors, buy up and sell down, causing disorder fluctuations and then spreading negative emotions such as fear across the market. Individual investors may follow and imitate institutional investors out of a belief that institutional investors possess greater capacity to collect and process information owing to professional knowledge. The key, however, is that many institutional investors are not rational, as assumed, and they are also susceptible to external shocks when dealing with information and making decisions. In addition, even for financial professionals, fear, a potential mechanism underlying risk aversion, might make investors divest more stocks \cite{fear_aer}. Then, fear from institutional investors might be magnified by following individual investors and reigniting much stronger disturbances that would lead to a market crash. Hence, from the perspective of negative emotions and their contagion, illiquidity can be contagious among stocks, suggesting that establishing a network by connecting stocks due to mutual illiquidity dependency could offer a new proxy for emotional contagion to finely investigate the dynamics of crashes.

Note that among the five emotion indicators, we only selected the indicator of fear due to its strongest and most significant correlation with illiquidity. The indicators of other emotions such as sadness, anger, joy and disgust, have weak and nonsignificant correlations with illiquidity and are accordingly not included in the present study.

\subsection{Illiquidity networks and crashes}
\label{subsec:inc}

In China's stock market, the actual interactions coupled within stocks are extraordinarily important because of susceptible investors. Most existing models forge links between stocks mainly based on the similarity of time series, e.g., price, and measures of Pearson and partial correlations are extensively employed \cite{MantegnaHierarchical,XuTopological,wang2018corr,KenettPartial}. However, the relationship between stocks is very complicated and should not be overly simplified to neglect trading behaviors, investor emotions and their possible contagions. Taking the limitations of linear correlations into account, we use mutual information to measure the nonlinear dependency between the illiquidity of stock pairs. The power to reflect the nonlinear dependency of mutual information in the networking market has been previously demonstrated and emphasized \cite{khoojineNetwork,BarbiNonlinear,MenezesOn}.

\begin{figure}[htp]
		\centering
		\includegraphics[width=120mm]{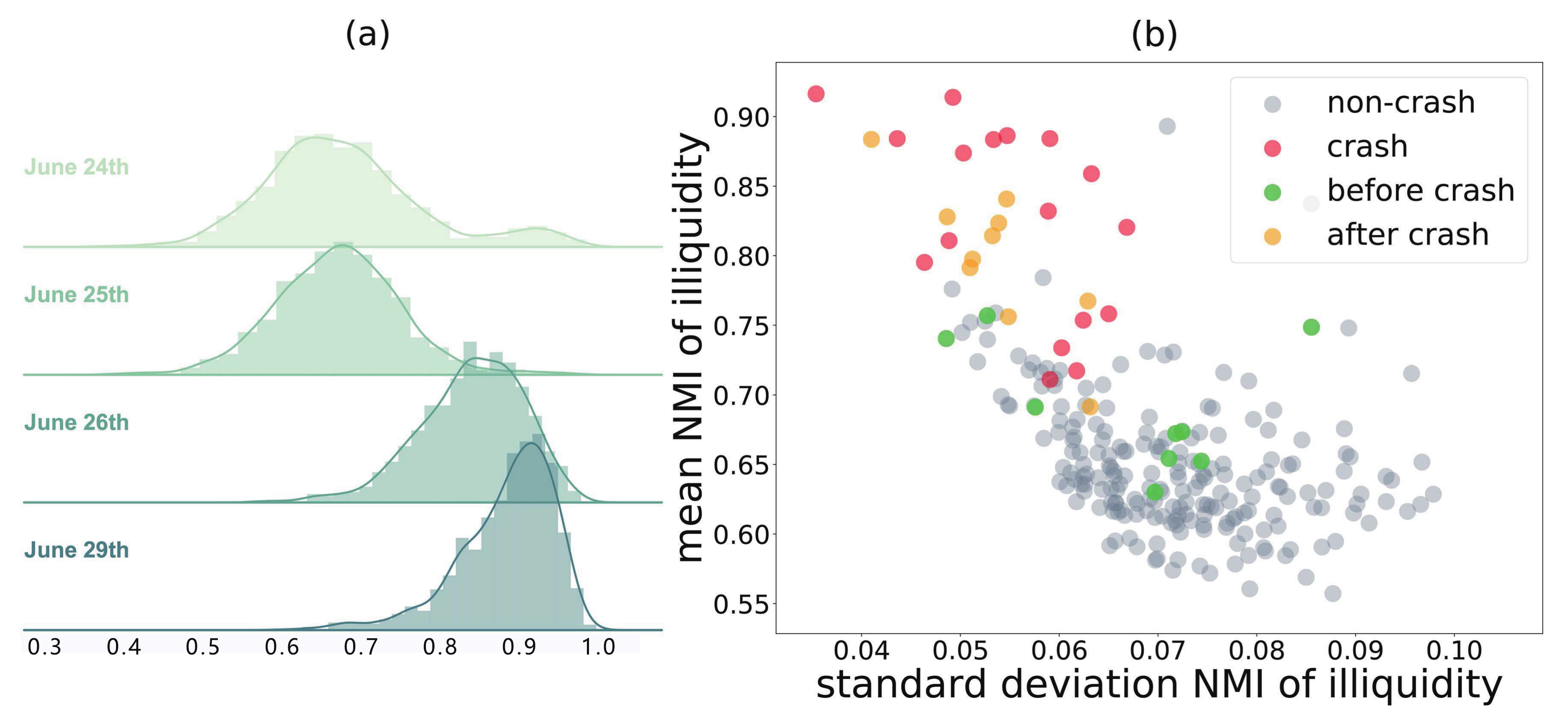}
		\caption{The normalized mutual information (NMI) of illiquidity (a) shows the distributions of NMI of illiquidity on both crash (June 26th and June 29th) and noncrash (June 24th and June 25th) days. It is clear that the globally averaged NMI increases while the standard deviation decreases when the stock market approaches turmoil. (b) shows the mean and standard deviation of the NMI of illiquidity with all transaction days in 2015.}
		\label{fig:nmi}
\end{figure}

By calculating the normalized mutual information (NMI) 
of illiquidity series in minutes between all pairs of stocks, we first attempt to profile the distributions of illiquidity dependency in the market on both crash and noncrash days. As can be found in randomly selected observations in Fig.~\ref{fig:nmi}(a), the globally averaged NMI increases while the standard deviation (e.g., the broadness of the distribution) decreases when the stock market approaches turmoil. Drawing a mean standard deviation graph with all the transaction days in 2015 for ease of observation (see Fig.~\ref{fig:nmi}(b)), it is clear that the average mutual information will increase and the standard deviation will decrease on crash days, indicating that the illiquidity network will become more closely connected and more homogeneously coupled when the market is in a bad situation. Due to pessimism, investors become cautious and unwilling to participate in transactions, which abruptly increases and spreads illiquidity across the market and results in a crash. In addition, we also find that market crashes exhibit a lasting effect because the days after the crash show the same characteristics as the day of the crash. However, regarding the days before the crash, as seen in Fig.~\ref{fig:nmi}(b), they overlap with those of noncrash days and hardly demonstrate any distinct features, suggesting that, from the global and static view, no warning signal can be detected. This inspires us to investigate the illiquidity network from more in-depth and dynamic perspectives.

\begin{figure}[htp]
	\centering
	\includegraphics[width=0.8\textwidth]{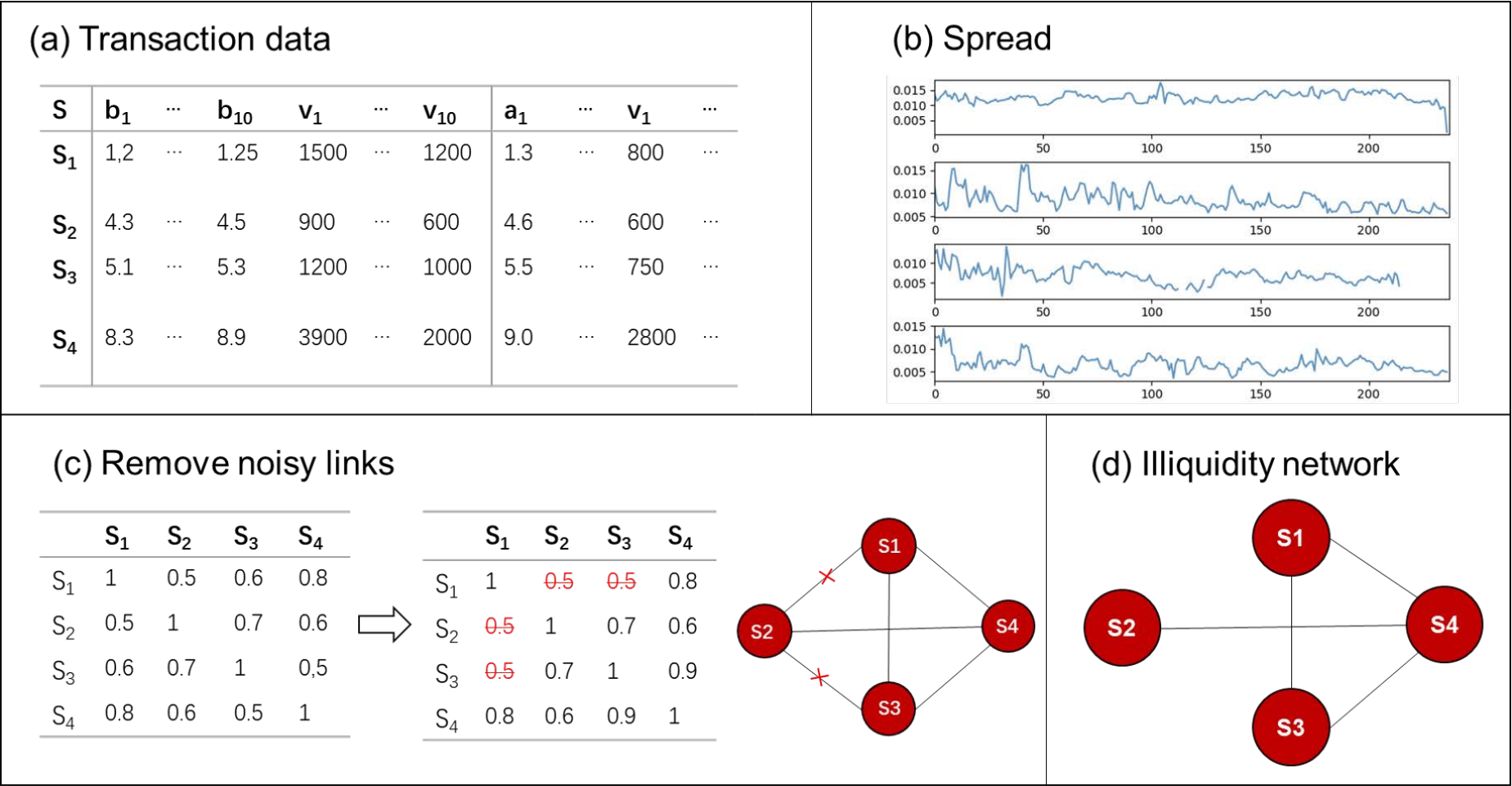} 
	\caption{The procedure for building the illiquidity network. (a) shows the raw data, including the bid price, ask price, bid volume and ask volume, which are divided into 10 levels. (b) shows the illiquidity we calculated through Equation~\ref{eq:ill}. (c) shows the removal of noisy links that might represent random dependency among stocks instead of plausible paths for illiquidity contagion. (d) shows a simple example of the established illiquidity network.}
	\label{network_construction}
\end{figure}

Figure~\ref{network_construction} illustrates the process of illiquidity network construction. Specifically, the entire construction procedure is primarily divided into the following steps: (1) we calculate the NMI of illiquidity series in minutes between all pairs of stocks; (2) we connect all pairs of stocks with undirected links, and these links are weighted by NMI; and (3) we remove links with weights lower than a threshold to refine and obtain the final network. Specifically, in building an illiquidity network, links are weighted as NMI between the illiquidity of their ends, while not all links are necessarily retained and those with lower weights, which might represent random dependency among stocks instead of plausible paths for illiquidity contagion, are removed.

\begin{figure}[htp]
		\centering
			\includegraphics[width=120mm]{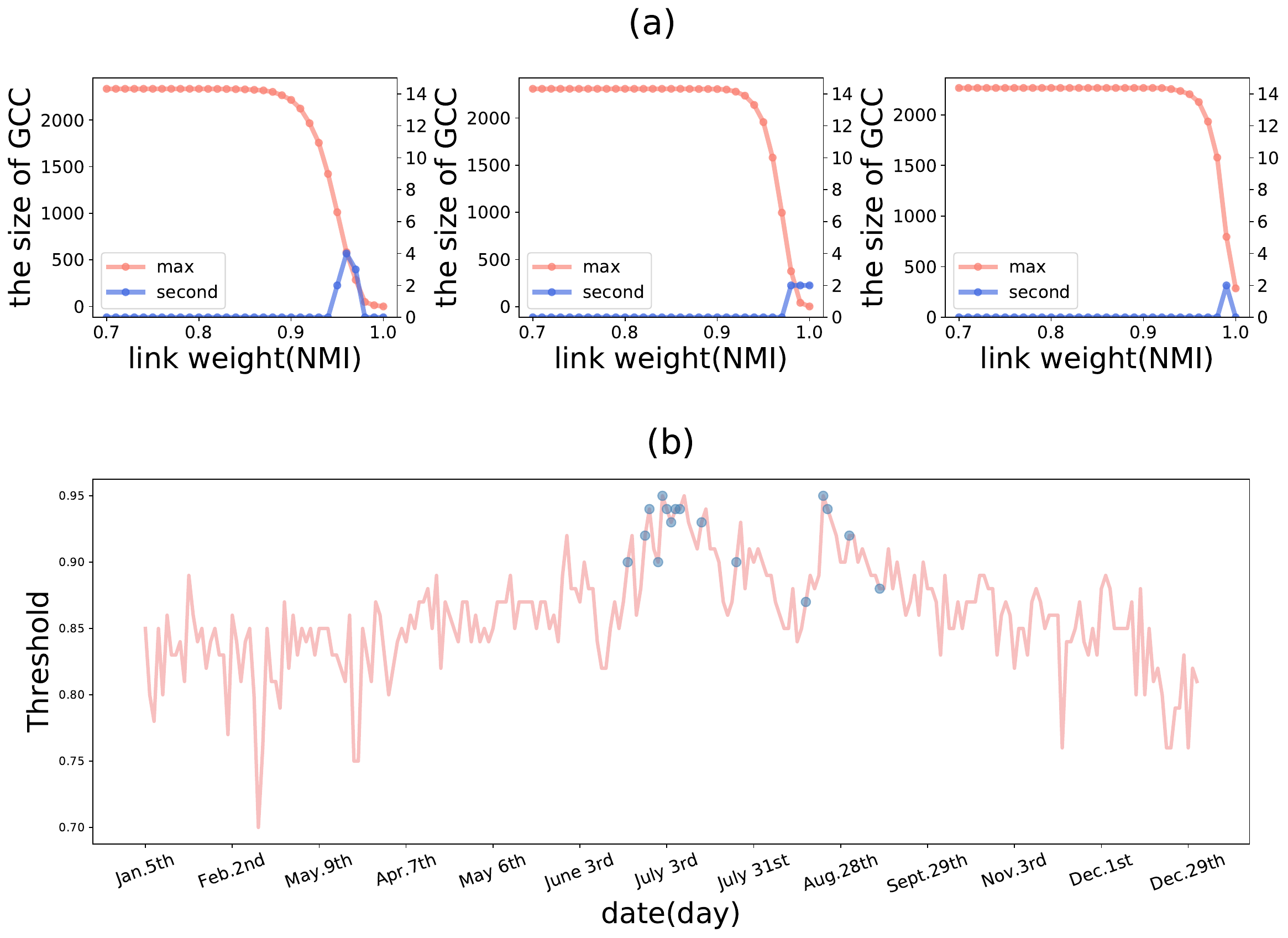}
			\caption{The threshold of link weights. (a) shows for three randomly selected trading days the sizes of the giant connected component (GCC) and the second-largest connected component vary as the threshold of link weights increases. The value beyond which the size of GCC starts to decline rapidly will be set as the threshold for each trading day, and it is found that the value can be well captured when the decline is more than 1\%. Considering that the size of the second-largest connected component is small, the GCC can well represent the entire network. (b) shows that the GCC ratio fluctuates over time but increases substantially during crash days (the red dots indicate crash days), consistently suggesting that the market will be more connected and coupled during crashes. }
		\label{fig:link_thr} 
	\end{figure}

Setting the threshold using values that are too small would result in dense but trivial networks, which would then undermine the possibility of detecting early warning signals of crashes. Setting the threshold with too high values will lead to sparse and even disconnected networks, which would eliminate the ability to model the market structure and profile crash dynamics. Similar to previous efforts~\cite{KenettDominating,LuHerding}, here, we present a way to target the appropriate value by observing how the giant connected component (GCC) of the illiquidity network decreases as the threshold increases. In particular, the size of the GCC is taken into account when locating the critical threshold of the link weight, i.e., the value beyond which the size of GCC starts to decline rapidly will be set as the threshold for each trading day, as seen in Fig.~\ref{fig:link_thr}(a). Links with weights below the threshold will then be omitted since their removal trivially influences the connectivity of the market structure. As in Fig.~\ref{fig:link_thr}(b), here, we set the threshold at 0.01, and other values both smaller and greater than it will be tested in later sections to demonstrate robustness. Fig.~\ref{fig:link_thr}(b) also shows that the ratio of the GCC in illiquidity networks fluctuates and significantly increases on crash days, consistently suggesting that the market will be more connected and coupled during crashes. High illiquidity dependency could facilitate the spread of illiquidity across the market, and a low deviation of illiquidity dependency would further lead to an abrupt collapse of the network. The positive associations between the GCC ratios of illiquidity networks and market crashes indicate that the structures refined by thresholds of link weights can serve as proper models of the  market network.

\begin{table}[htp]
		\caption{Sectors and capital styles of stocks}
		\label{tab:ss}
		\begin{tabular}{lllll}
			\cline{1-2}
			\multicolumn{1}{c|}{sector} & \multicolumn{1}{c}{\begin{tabular}[c]{@{}c@{}}Agriculture, Communication and cultural,Comprehensive,Construction,\\Electricity, Gas, Water, Extractive, Financial, Information technology,\\Manufacturing, Real estate, Retailing, Service, Transportation\end{tabular}} &  &  &  \\ 
			\cline{1-2}
			\multicolumn{1}{c|}{style}  & \multicolumn{1}{c}{\begin{tabular}[c]{@{}c@{}}Small-cap-growth, Small-cap-balance, Small-cap-value\\ Mid-cap-growth, Mid-cap-balance, Mid- cap-value\\ Large-cap-growth, Large- cap-balance, Large- cap-value\end{tabular}} &  &  &  \\ 
			\cline{1-2}
			&  &  &  &  \\ &  &  &  & 
		\end{tabular}
	\end{table}

\begin{figure}[htp]
		\centering
		\includegraphics[width=120mm]{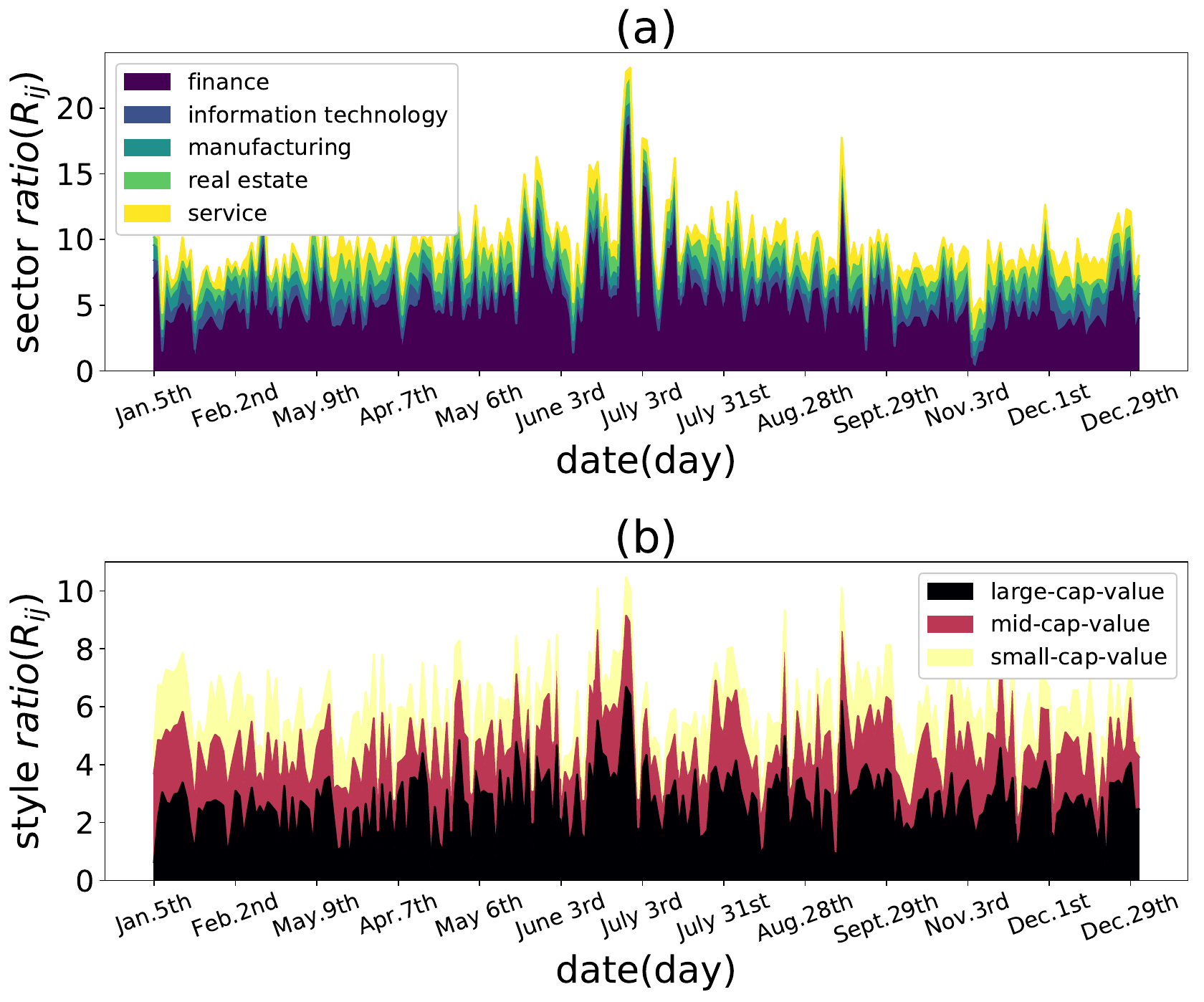}
		\caption{The degree-weighted shares of different sectors and capital styles in illiquidity networks. (a) shows the proportions in different sectors. Note that the proportions in other sectors are very similar except for finance, and therefore, only a few representative industries are selected to simplify the picture. (b) shows the proportions of different capital styles of stock values; the large-cap-value is the most critical group in market crashes. For growth stocks and balanced stocks, the results are the same, and thus they are not shown in the figure.}
	\label{fig:opro}
\end{figure}

The illiquidity network of the stock market evolves in the form of adding new links or removing existing connections. We find that China's stock market evolves at a high frequency, especially on crash days, and only 10\% of links remain on average for two consecutive trading days (see Appendix Fig.~\ref{fig:link}). Highly varying structures suggest that targeting critical stocks with a profound role in crashes can help assess market risk. In terms of grouping stocks into different sectors or capital styles (see Table~\ref{tab:ss}), a degree-weighted proportion, denoted as $R_{ij}$, is defined to identify key group $i$ of stocks on trading day $j$. Specifically,
\begin{equation}
\label{eq:Rij}
{R_{ij}} = \frac{n_{ij}/n_j}{N_{ij}/N_j},
\end{equation}
where $n_{ij}$ is the occurrence of stocks belonging to group (sector or style) $i$ and is summed over all links in the network on day $j$, $n_j$ is the occurrence of all stocks and is summed over all links in the network on day $j$, $N_{ij}$ is the number of group $i$ in the network on day $j$, and $N_j$ is the number of unique stocks in the network on day $j$. Accordingly, the group of stocks with higher $R_{ij}$ will occupy more links in the market, meaning that they are more dependent on other stocks' illiquidity and have a greater likelihood of accumulating or passing on crash risk. Unexpectedly, as shown in Fig.~\ref{fig:opro}(a), the finance sector constantly occupies the highest proportion in China's stock market, especially on crash days. Regarding the capital style, the style of large capitalization, i.e., the large-cap-value, is the most critical group in market crashes (see Fig.~\ref{fig:opro}(b)). Both observations suggest that stocks in finance, especially those with large capital values, should be inspection targets for market regulators.

\begin{figure}[htp]
		\centering
		\includegraphics[width=120mm]{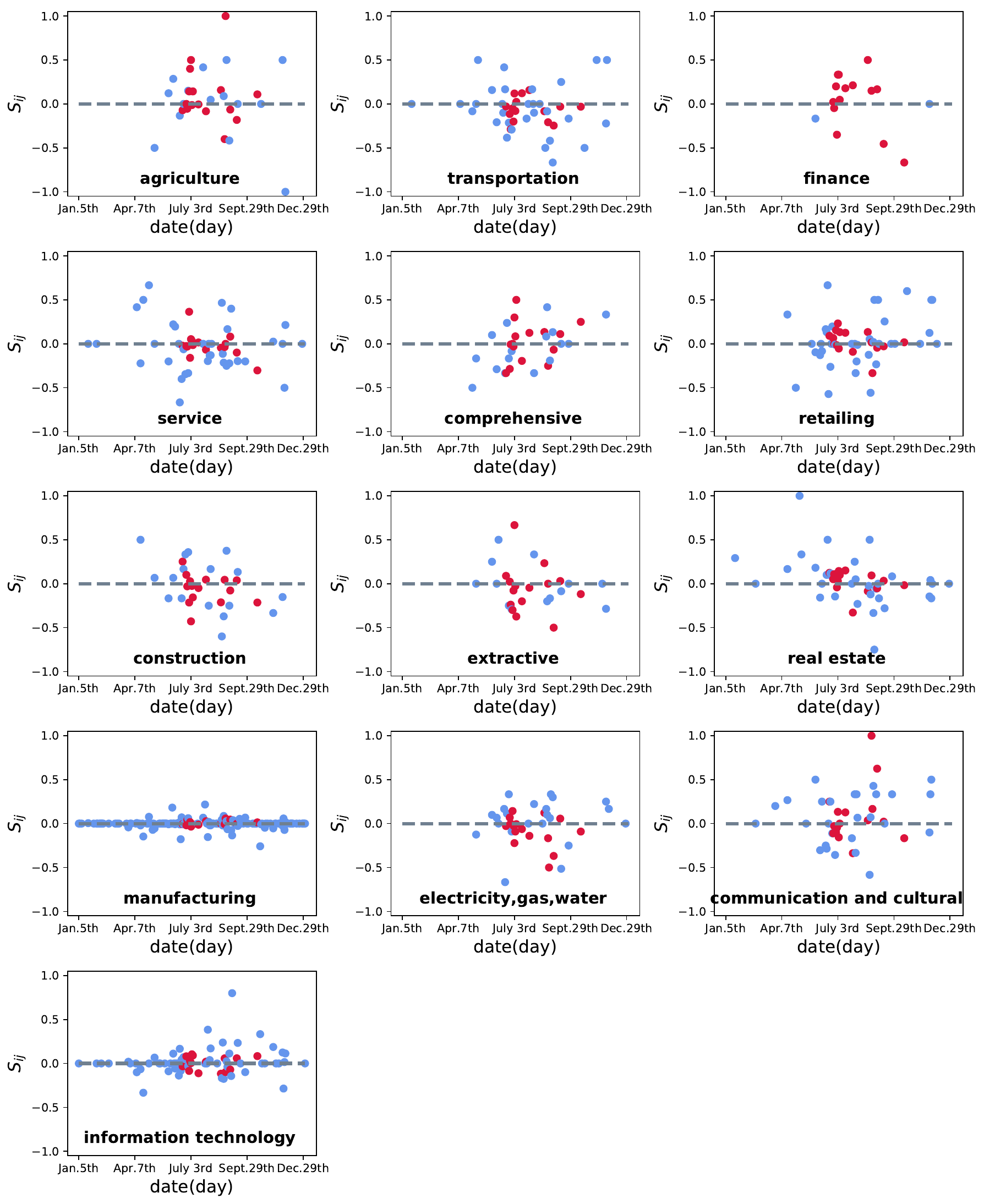}
		\caption{The significance of failing before the peak. The red dots indicate crash days, and the blue dots indicate noncrash days. $S_{ij}=R_{ji}^{bp}- R_{ji}^{bpr}$, so $S_{ij}$ may be positive or negative. If $S_{ij}$ is positive, then stocks within $i$ tend to fail before peaks. It is obvious that the finance sector fell the most before peaks on crash days. In contrast, sectors such as manufacturing and information technology perform similarly on both crash and noncrash days. Note that $S_{ij}$ cannot be calculated for all stocks since some of them might not appear in the illiquidity network due to good liquidity, especially on noncrash days.}
		\label{fig:peaks}
\end{figure}

The collapse of China's market during the crash consisted of waves of stocks completely losing illiquidity, i.e., declining to the down limit~\cite{lusmall}. These failure waves produced peaks in the number of newly failed stocks (see the example in Appendix Fig.~\ref{fig:peak2}). Assuming that each wave of failure can be identified by a peak, then stocks that failed before the peak could be seeding failures that lead to the corresponding wave of illiquidity loss. Then, sectors with more stocks that failed before peaks might be the causes of the subsequent collapse and thus could be targets for early inspection and even sources of warning signals. A new ratio, denoted as $R_{ij}^{bp}$, is thus defined to target critical sectors, which can be calculated as $$R_{ij}^{bp}=\frac{n_{ij}^{bp}/N_j^{bp}}{N_{ij}/N_j},$$ where $n_{ij}^{bp}$ is the number of stocks that failed before peaks in group $i$ on day $j$ and $N_j^{bp}$ is the number of stocks that failed before peaks on day $j$. To test the significance of failing before peaks, the failure time for all the stocks on one trading day are also randomly shuffled to obtain a random value of $R_{ij}^{bp}$, which is denoted as $R_{ij}^{bpr}$ for comparison to test significance. Then, for group $i$, the probability of being seeds that lead to a wave of failures on day $j$ can be defined simply as $S_{ij}=R_{ij}^{bp}- R_{ij}^{bpr}$. 
Intuitively, $S_{ij}$ will be much greater than 0 if stocks within $i$ tend to fail before peaks. Consistent with our above observation, the finance sector, as seen in Fig.~\ref{fig:peaks}, failed the most before peaks, especially on crash days. In contrast, the significance of sectors such as manufacturing and information technology fluctuates around zero with trivial deviations. This again suggests that finance stocks in China's market might be sinks or even triggers that produce illiquidity and spread it across the market. In terms of inspecting these finance stocks, market practitioners, particularly regulators, could provide warnings based on their abnormal variations in illiquidity. In fact, the finance-related stocks' occupation of the most central positions within the illiquidity network is associated with exclusive investment styles instead of their relative sizes in Chinese stock market, i.e, stocks in the finance sector are repeatedly preferred in the portfolios of different investors. Direct and consistent evidence has been also provided by Lu et al. (2018)~\cite{LuHerding} that `too-connected' stocks that resulted from herding in portfolios mainly belong to the finance sector and most of them are large-cap-value stocks~\cite{LuHerding}.

It is expected that to identify key transmitters, centrality measures rather than degrees, in particular betweenness centrality derived from illiquidity networks, could offer more straightforward and even sophisticated approaches for regulators. Specifically, as Equation~\ref{eq:Rij} defines, we replace degree with betweenness to weight the relative proportions of every sector and capital style in the illiquidity networks. As seen in Appendix Fig.~\ref{betweenness_ratio}, for neither the section ratio nor the style ratio do stocks from the finance sector or with large market capital demonstrate distinct patterns over time. We also test the significance of failing before peaks, as shown in Appendix Fig.~\ref{betweenness_sector_ratio}, and there are no significant sectors that tend to fail before peaks when weighted by betweenness. Therefore, degrees in the illiquidity network unexpectedly reflect stocks' positions in market crashes, and they will be investigated further in the profiling of crash dynamics.

\begin{figure}[htp]
		\centering
		\includegraphics[width=120mm]{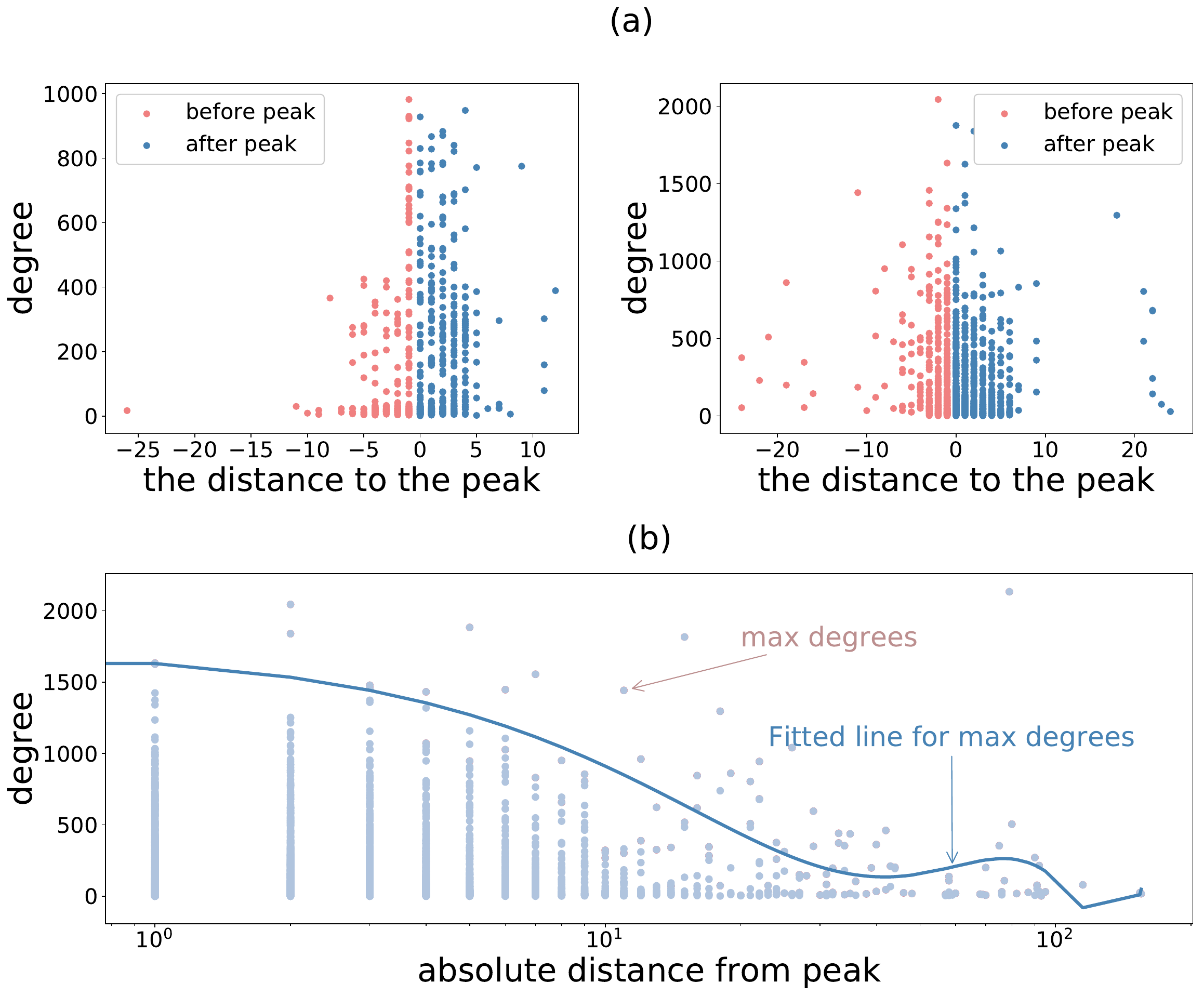}
		\caption{The correlation between stock degree and the time from losing illiquidity to the peak. (a) shows the degrees of stocks that decline to the down limit before and after the peak. Peaks stands for minutes in which the number of failed stocks hits the local maximum~\cite{lusmall}, indicating a wave of stocks at the down limit and might result in a market crash. How to identify and determine the peak of stock declines to the down limit is illustrated and demonstrated in Appendix Fig.~\ref{fig:peak2}. (b) shows that the greater the absolute distance is, the smaller the degrees of stocks (the y-axis is logarithmic). Considering the considerable divergence of degrees in each bin of distance, we fit the correlation between the maximum degree of each bin and the distance (with the coefficient being -0.66 and the p-value being 0.00) through a polynomial fit of least squares (numpy.polyfit in Python) into a solid line to further and vividly demonstrate this association. }
		\label{fig:tdistance}
\end{figure}

The illiquidity network can also track the dynamics of market crashes. Considering that peaks of newly failed stocks can be interfaces to split failure cascades, the time between the loss of illiquidity to the peak inherently measures the stage at which the stock joins the crash cascade. Specifically, for negative distances, smaller distances represent an early collapse, while for positive distances, greater distances represent later failures in the crash (see Fig.~\ref{fig:tdistance}(a)). We then examine the function between stock degree and the absolute value of time distance, as shown in Fig.~\ref{fig:tdistance}(b). It is found that the degree, in particular the maximum degree in each bin, is negatively correlated with the distance. This negative association indicates that stocks that fail at nearly the peak are those with high degrees, while those that fail at the early state or at the end of the crash possess small degrees. That is, the crash starts from stocks with small degrees and then spreads to stocks with high degrees, which are usually located at the core of the network, and finally cascades to the periphery. Although market crashes in essence originate from the failure of these crucial nodes in the core, those with small degrees of collapse at the early stage might be the real triggers. Consistent with a previous study~\cite{lusmall}, this finding discloses the unexpected role of small-degree stocks in market crashes and suggests that regulators pay more attention to those that might conventionally be overlooked, especially those in the finance sector. Stocks with fewer connections, compared to those with more connections, might be more sensitive to interior or external perturbations, thus demonstrating a greater likelihood of losing liquidity and  triggering a panic. Then, their illiquidity, in particular that conveyed by panic, would be propagated among investors, amplified and eventually result in illiquidity of more strongly connected stocks that is then broadcast to more stocks.

\subsection{Illiquidity networks and a warning signal}
\label{subsec:wsignal}

\begin{figure}[htp]
			\centering
			\includegraphics[width=120mm]{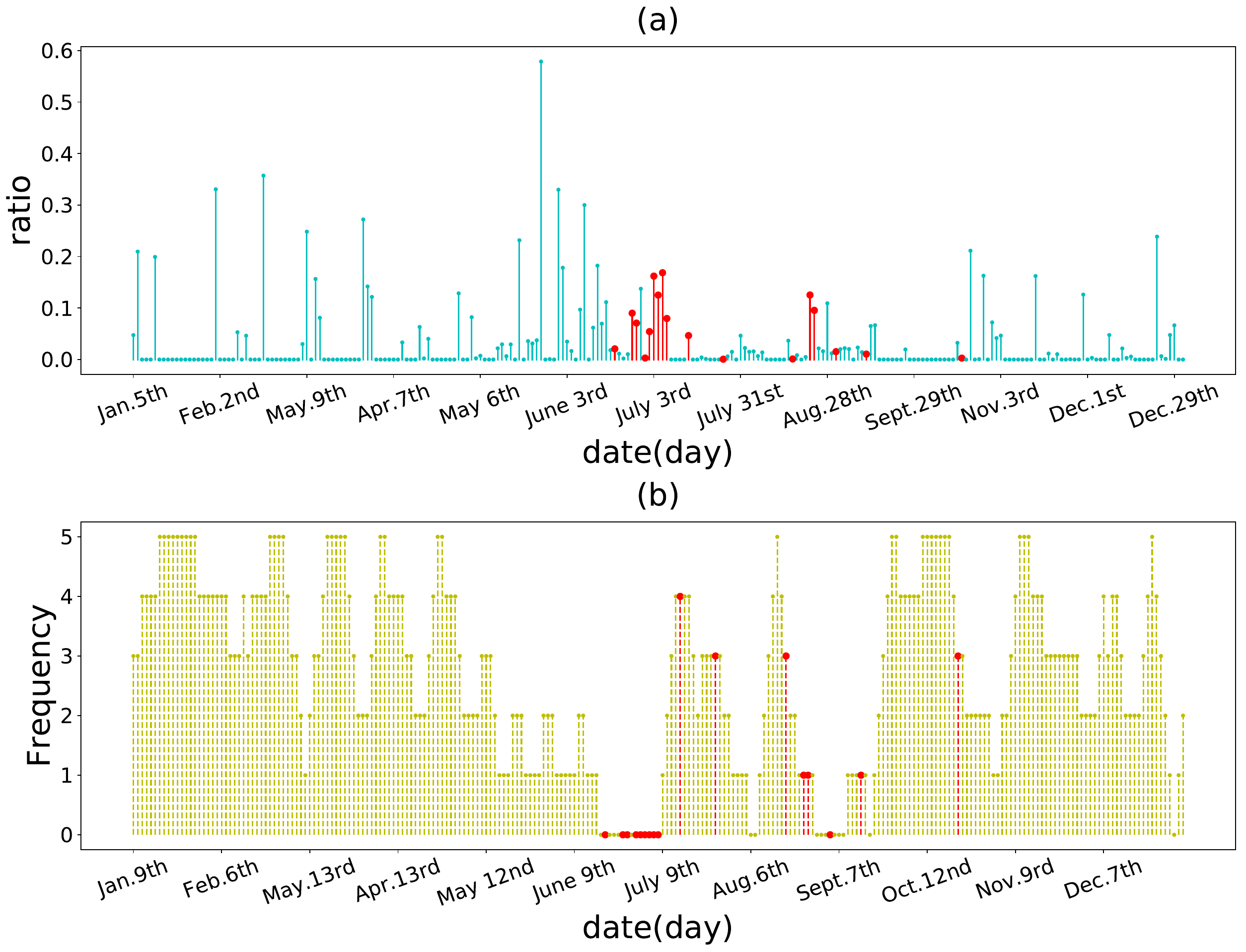}
			\caption{The warning signal. (a) shows the likelihood of systemic failures instead of random errors. The red bar indicates crash days, and the blue bar indicates noncrash days. (b) shows the occurrence of $w_d=0$, i.e., the daily nonrandomness is zero within five days, which is denoted as $N_{w_d=0}$ to construct a warning signal. Specifically, a smaller $N_{w_d=0}$ suggests more systemic failures and a greater likelihood of leading to a market crash. An abrupt decline in $N_{w_d=0}$ can be detected one day earlier. This indicates that if $N_{w_d=0}=0$ in the previous five days, a warning signal can be implemented to warn of a market crash the next day. }
			\label{fig:warn} 
	\end{figure}

The above illustrations solidly suggest associations between illiquidity networks and market crashes. Assuming that market crashes are systemic failures rather than random errors, stocks that failed jointly in a short interval, e.g., ten minutes, should be inherently entangled with each other due the contagion of losing illiquidity and therefore connected in our illiquidity network. Then, the nonrandomness of failures within a short interval $i$ can be defined as $w_i=\frac{e_{n_f}}{n_f(n_f-1)/2}$, where $n_f$ stocks simultaneously reached the down limit at $i$, $e_{n_f}$ is the number of links among them captured in the illiquidity network built on the corresponding day and $n_f(n_f-1)/2$ is the maximum number of possible links among them. In line with this, a higher $w_i$ represents a greater likelihood of systemic failures instead of random errors, i.e., signs of a crash. $wd_j=\textless w_i\textgreater$ from all intervals of trading day $j$ can be accordingly measured to value the daily nonrandomness. As seen in Fig.~\ref{fig:warn}(a), most values of the daily nonrandomness are zero, and greater fluctuations significantly occur as crash days approach, which implies that a warning signal could be accordingly forged.

\begin{figure}[htp]
		\centering
		\includegraphics[width=120mm]{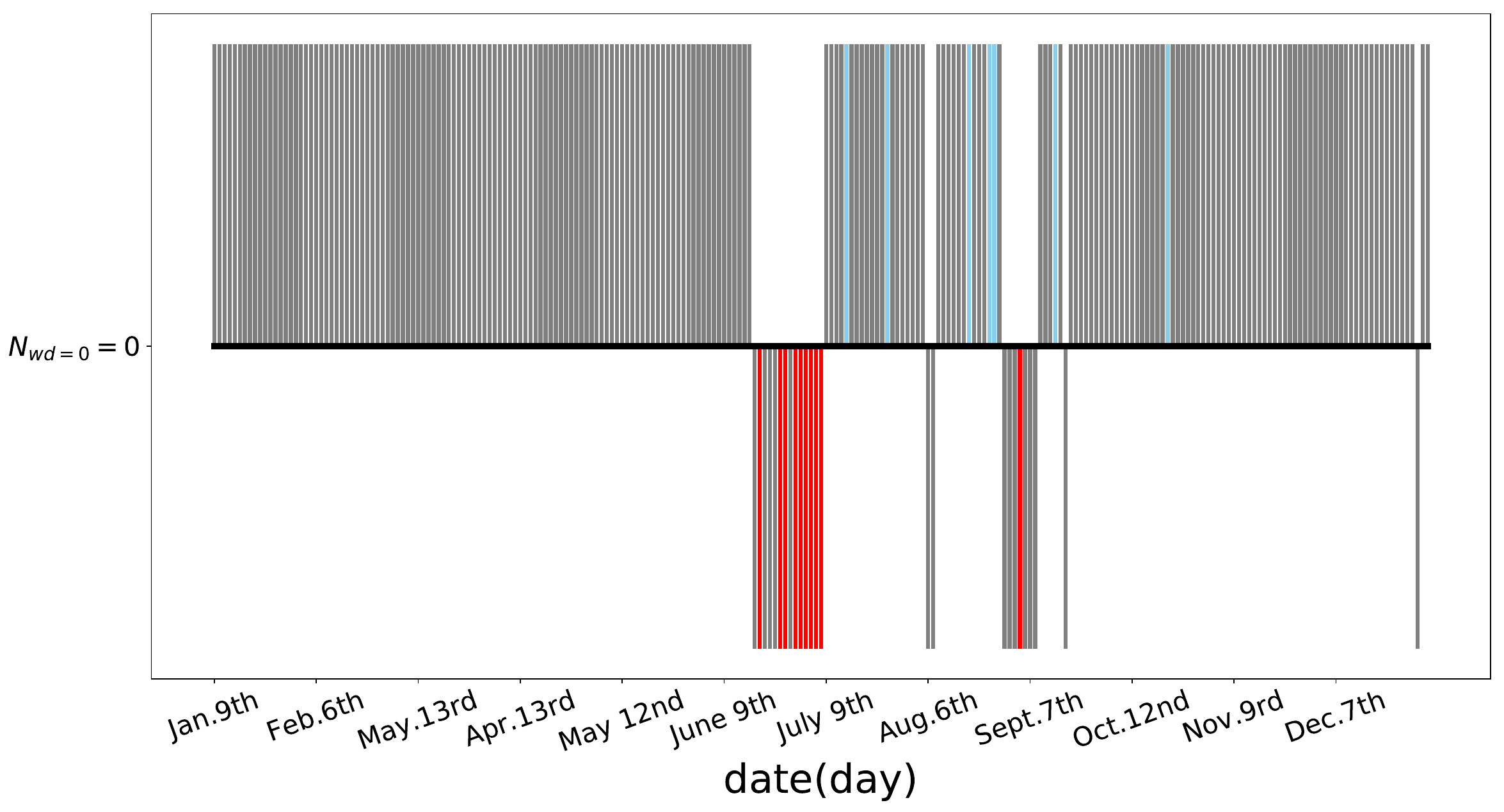}
		\caption{The warning signal of $N_{w_d=0}=0.$ Gray bars stands for noncrash days, red bars mark crash days that can be warned of one day in advance, and those that cannot be warned of in advance are colored blue.}
		\label{fig:signal}
\end{figure}

Given the fluctuations of daily nonrandomness (see Fig.~\ref{fig:warn}(a)), a sliding window of $t$ days, meaning that historical information of the previous $t$ days is expected to be helpful, is set to smooth the daily sequence, and we then simply count the occurrences of $w_d=0$ within the window, which is denoted as $N_{w_d=0}$ to construct a warning signal. Specifically, a smaller $N_{w_d=0}$ suggests more systemic failures and a greater likelihood of leading to market crash. As seen in Fig.~\ref{fig:warn}(b), as $t=5$, an abrupt decline in $N_{w_d=0}$ can be detected one day earlier than more than half of the 2015 crash days in China's stock market, particularly for those consecutive crash days that occurred in the early phase. This indicates that if $N_{w_d=0}=0$ in the previous five days, a warning signal should be sent out because there will be a market crash the next day, as seen in Fig.~\ref{fig:signal}. Note that $t=5$ is the optimal setting, as we vary $t$ from 1 to 15 days. Interestingly, time windows with lengths shorter than five days result in insensitive $N_{w_d=0}$, while those longer than five days result in the disappearance of signals in advance. One day ahead of the crash is vital because it indicates that the presented early warning signal can help prevent systematic risk of the market in reality. Note that not all the crashes in 2015 can be effectively and correctly warned of (see Fig.~\ref{fig:warn}(b) and Fig.~\ref{fig:signal}) and those for which the signal failed might be caused by shocks that are similar to random ones on noncrash days. In addition, past crashes might essentially restructure the stock market and make later crashes more difficult to predict.

\section{Robustness test}
\label{sec:robust}

In establishing the illiquidity network, noisy links caused by random associations are omitted by a threshold determined by the variation in the GCC. To check the robustness of our results regarding this critical parameter, more threshold settings are tested. In particular, the ability to detect crashes through the established early signal is evaluated on additional values other than 0.01. As expected an can be seen in Appendix Fig.~\ref{border_zero_bar}, there is a threshold interval within which the results of detecting crashes in early days are stable. On the one hand, the early warning signal is ineffective when the threshold is less than 0.007 (see Appendix Fig.~\ref{border_zero_bar}(a)), which becomes too sensitive and misrecognizes the noncrash day as the crash day. On the other hand, the early warning signal is ineffective and misjudges a sharp market crash in June when the threshold is more than 0.02 (see Appendix Fig.~\ref{border_zero_bar}(d)). Therefore, a threshold in the range of (0.006,0.03) can be set to obtain a good signal, and the setting of 0.01 that we adopted in this study is proper.

\begin{figure}[!htbp]
	\centering
	\includegraphics[width=1.\textwidth]{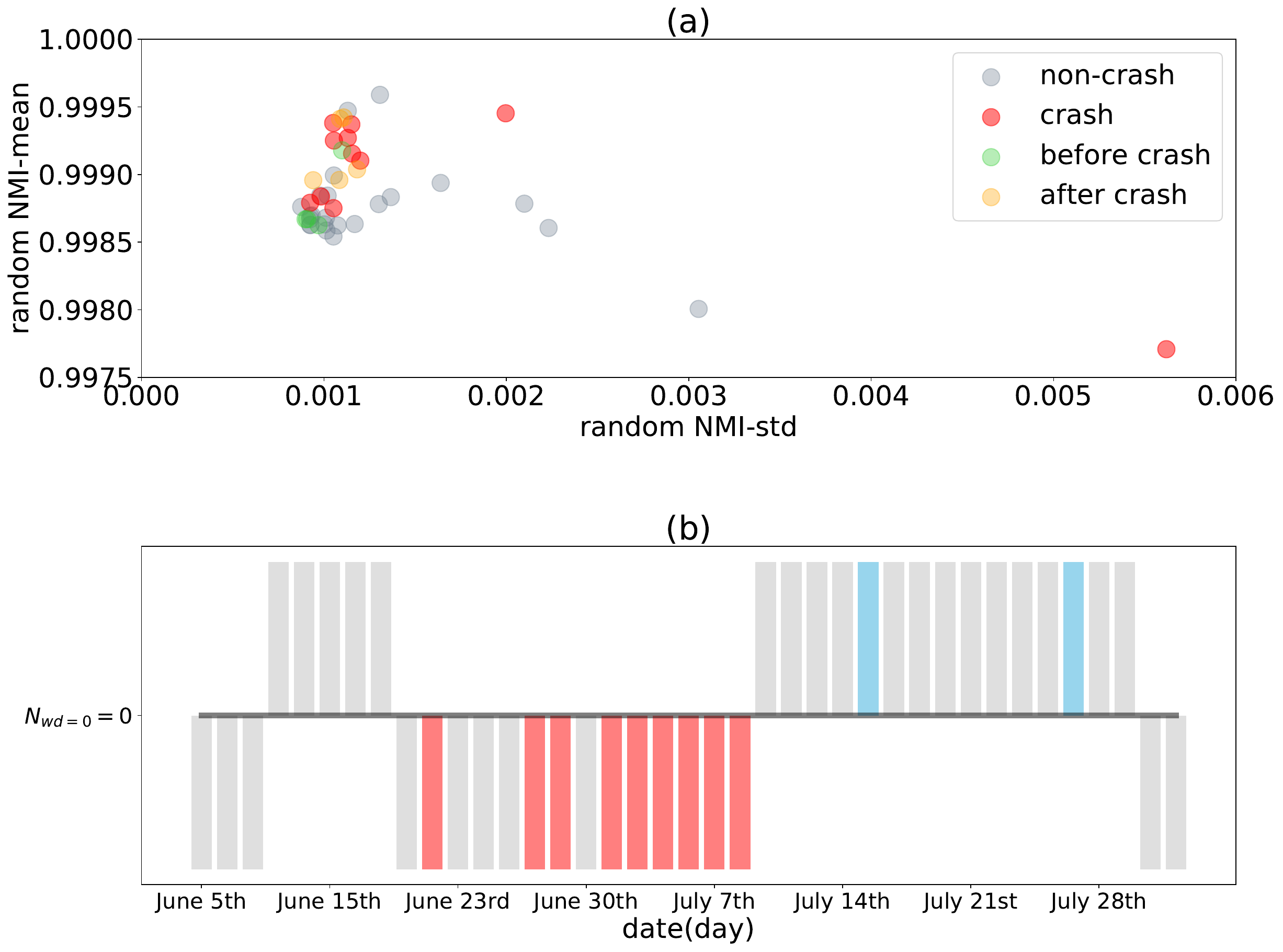} 
	\caption{ Results after randomization of quotation data. Here, we only test the robustness in two key months of June and July, when market crashes occurred frequently. (a) shows the mean and standard deviation of random NMI, and there is no clear boundary between the crash and the noncrash days. (b) shows the early warning signal after randomization. Specifically, gray bars represent noncrash days, red bars mark crash days that can be warned of one day in advance, and those that cannot be warned of in advance are colored blue. However, the most catastrophic crashes on 26 June and 29 June are missed. }
	\label{random_NMI}
\end{figure}

It is also possible that our results, in particular the difference between crash and noncrash days, could not be significantly distinguished from that of randomly connected illiquidity networks. To test this, we construct random illiquidity networks by separately shuffling the original order of both the ask and bid sequences. The calculation of the spread, the generation of the illiquidity sequence, the calculation of the NMI between stock pairs and the removal of links with weights lower than the threshold (i.e., 0.01) are similarly performed without changing any parameters involved. As shown in Fig.~\ref{random_NMI}, it is clear that after randomization, the NMI between stock pairs cannot capture the difference between crash and noncrash days, and the early warning signal is ineffective because it fails to warn of the most catastrophic crashes. These results confirm that our findings on illiquidity networks are robust and significant. The comovement of illiquidity series captured by the NMI is not spurious but carries instructive information for warning of crashes in early stages.

\begin{figure}[htp]
	\centering
	\includegraphics[width=1.\textwidth]{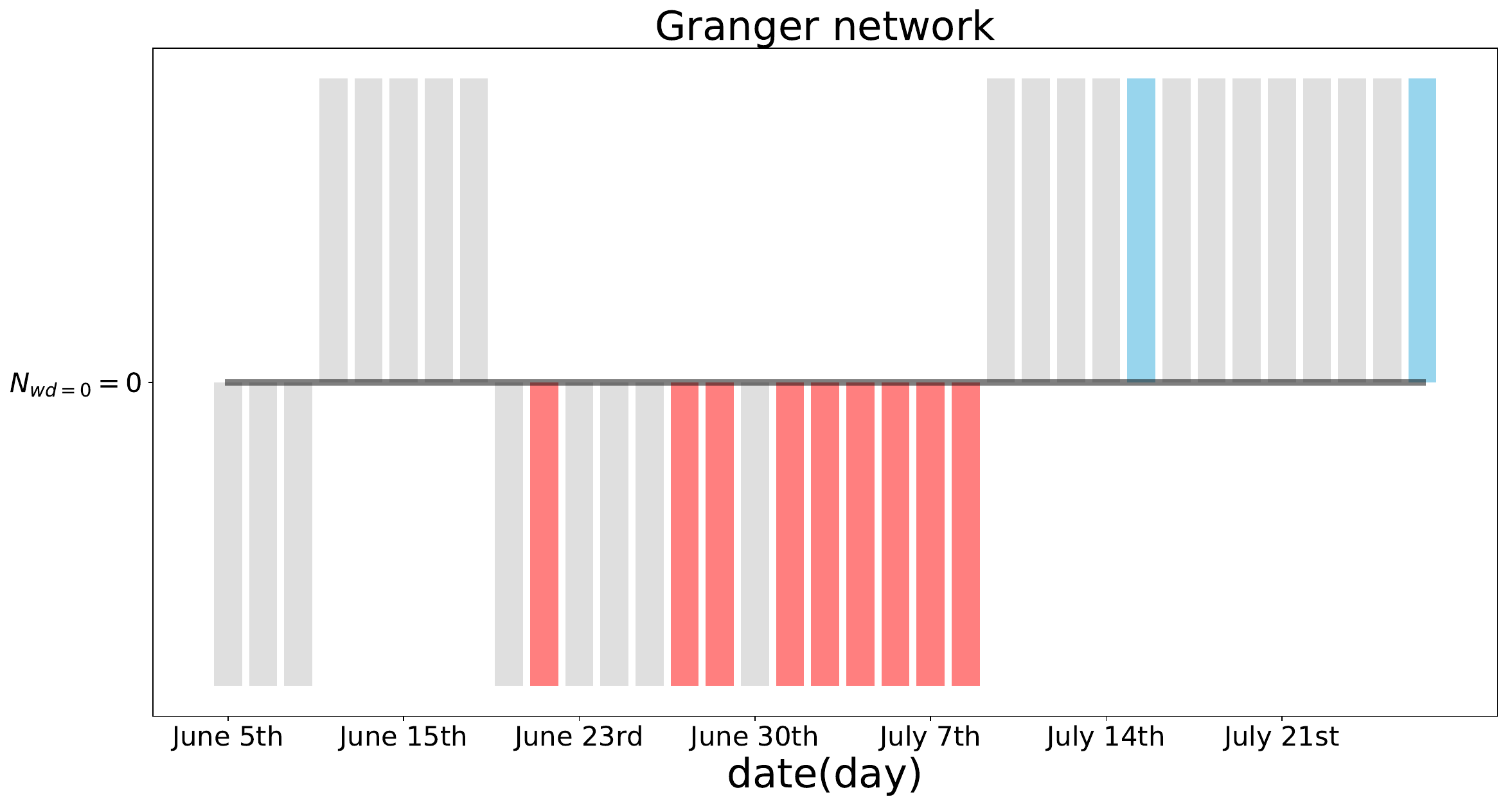} 
	\caption{The early warning signal of Granger networks (time-lag=1, risk level=0.05). Gray bars represent noncrash days, red bars mark crash days that can be warned of one day in advance, and those that cannot be warned in advance are colored blue. Here, we adopt only the two key months of June and July (when market crashes occurred frequently) due to the long calculation cycle.}
	\label{granger_zero_bar}
\end{figure}

Stock networks can also be forged through linear correlations and the Granger test, which takes causality into account. A natural question is whether networks established in these traditional manners could be employed to understand market crashes and even detect early signals of a crash. Hence, we further test the robustness of our results in these traditional networks. We first build illiquidity networks in terms of Pearson correlations between illiquidity sequences of all stock pairs. As shown in Appendix Fig. \ref{pearson_mean_std}, however, crash days cannot be distinguished from noncrash days through these Pearson correlation networks. This indicates that the networks derived from the linear correlation of illiquidity series lose the capability of warning of crashes, and the use of nonlinear measures such as NMI in building illiquidity networks is crucial. We further establish Granger networks by calculating the Granger coefficients of illiquidity series between all pairs of stocks. Considering the configuration in which the time lag is 1 and the confidence level is 95\%, as shown in Appendix Fig.~\ref{granger_edges}, the results are similar to those of the NMI illiquidity network. Specifically, the manifested market structure also evolves at a very high frequency, in which only 10\% of links are sustained for two consecutive trading days. Note that the early warning signal is also consistent, as shown in Fig.~\ref{granger_zero_bar}, which again strengthens the credibility of our results derived from the NMI. However, these unchanged links remain unclear, as seen in Appendix Fig.~\ref{removing_links}. We also construct Granger networks with various settings of a time lag of 2, confidence level of 95\% and a time lag of 1, confidence level of 99\%. The early warning signal becomes sensitive in both configurations, and the warning effect is missing. That is, similar to illiquidity networks based on the NMI, key parameters in constructing networks are indeed associated with the ability to warn of crashes in the early stage. Compared to the NMI, although consumers would require much more intensive calculations, constructing Granger networks does not generate new power to explain the unchanged links.

In fact, building the illiquidity network differs from the construction of conventional financial networks in two respects. First, illiquidity can be effectively measured at a fine resolution and, thus, exclusively represents the minimum decision unit in stock investments. For existing approaches based on price correlations and Granger causality, time series of too high frequency might lead to trivial sequences with no obvious trend, and some local changes may even result in biased associations, e.g., prices might remain stable in short periods, and abrupt changes in certain times disturb the calculation of correlations. Second, illiquidity is found to be significantly associated with the emotions of investors, particularly fear. Thus, the established illiquidity network is inherently coupled with investors' emotions and can offer insightful observations in understanding market crashes. It is challenging for alternative approaches based on correlations to directly and explicitly account for behavioral factors (e.g., emotions and decisions).

\section{Conclusions and limitations}
\label{sec:con}

Financial systems such as stock markets are vital components of modern economies and play profound roles in economic growth. Market crashes, however, occur occasionally and generate very large shocks to the entire socioeconomic system and even lead to global recessions. For example, the crash in China's stock market in 2015 erupted unexpectedly, and over one-third of the market value abruptly evaporated. How to understand and warn of crashes has been an open and trending problem not only in finance but also from interdisciplinary perspectives. In fact, from the perspective of system science, the stock market can be modeled as a complex network, and crashes can thus be cascading failures of stocks that decline to the down limit. Nevertheless, in previous studies, trading behaviors, in particular the emotions of investors, have rarely been considered in networking the market, which in essence motivated the present study.

As an empirical study seeking to profile the market structures underlying crashes and present early warning signals, we mainly pursued a data-driven paradigm. However, economic intuitions and behavioral hypotheses are indispensable aspects of our investigation. Regarding economic intuition, it is assumed that different stocks could be essentially associated because of common investors, in particular individual investors. Regarding behavioral hypotheses, given the domination of individual investors in transactions, we assume that in China's stock market, investor behaviors, in particular irrational behavior, could help disseminate disturbances, amplify panic and even produce systemic crashes. Their abnormal decisions and negative emotions could help profile and even warn of market crashes. It is further anticipated that indicators of market crashes could be identified through high-frequency transactions, which record the elementary decision units of all investors and thus potentially represent detailed investment behaviors. Based on these intuitions and assumptions, illiquidity is selected to construct the market structure between stocks because it can be derived from high-frequency ask-bid prices and demonstrates strong associations with the fear of investors that represents panic.

By connecting stocks with mutually associated illiquidity, we find that the market is more densely and homogeneously connected due to a large mean and low deviation of illiquidity dependencies, which can describe the abrupt collapse of the market during a crash. Stocks are not randomly connected, and those with large capital value or that are from the finance sector are targeted as the most influential stocks in a market crash. What is even more interesting is that the negative correlation between the maximum degrees and distances to the peak of the lower limit suggest a pattern of periphery-core-periphery propagation in crashes. By simply counting the days without systemic failures in the previous five days, an early signal is also derived from the illiquidity network that can warn of more than half of the crash days in 2015. Its robustness is further tested by varying parameters and considering different network construction methods. Our results could help market practitioners such as regulators inspect risky stocks such as those from the finance sector or with small degrees and identify early warning signals to prevent crashes.

This study is limited in establishing emotion indicators due to the extreme scarcity of stock-related tweets on Weibo. Nevertheless, given the significant association between the fear index and market illiquidity, it is argued that the presented high-frequency illiquidity could, to some extent, reflect the possible comovement of emotions between stocks. However, exploring approaches to emotion measurement from high-frequency records instead of social media is a necessary and promising direction.

Meanwhile, the unchanged links that exist across two consecutive trading days in illiquidity networks remain unclear, even in networks constructed using the Granger causality of illiquidity. We only confirm that these links do not help keep the market structure connected and facilitate the detection of warning signals. However, what causes these sustained links demands further investigation in future work.

We also recognize that not all crashes can be warned of accurately by the proposed early signal ($N_{w_d=0}=0$). Those crashes that our signal failed to warn of imply that the causes of crashes can be sophisticated and that some of them might truly be caused by random shocks. How to group crashes into categories of those that can be warned of and those that cannot will be an interesting direction in future work. In the meantime, the possible entanglement between different crashes also deserves further effort.

Our economic intuitions and behavioral hypotheses are established in the context of China's stock market, which possesses a large volume of individual investors that represent most transactions. Hence, we are not confident that our methods of establishing illiquidity networks and detecting early signals of crashes could be easily extended to other markets. Because of this, it could be a promising direction in future work to test our approaches in more markets.


\section*{Conflict of Interest}
The authors declare that they have no conflicts of interest.

\section*{Acknowledgements}
This research was supported by National Natural Science Foundation of China (Grant No. 71871006). The authors also appreciate valuable comments and constructive suggestions from Dr. Shan Lu.

\begin{appendices}
	\section*{Appendix}

	\setcounter{figure}{0}
	\renewcommand{\thefigure}{A\arabic{figure}}

	\begin{figure}[htp]
  	\centering
  	\includegraphics[width=.8\textwidth]{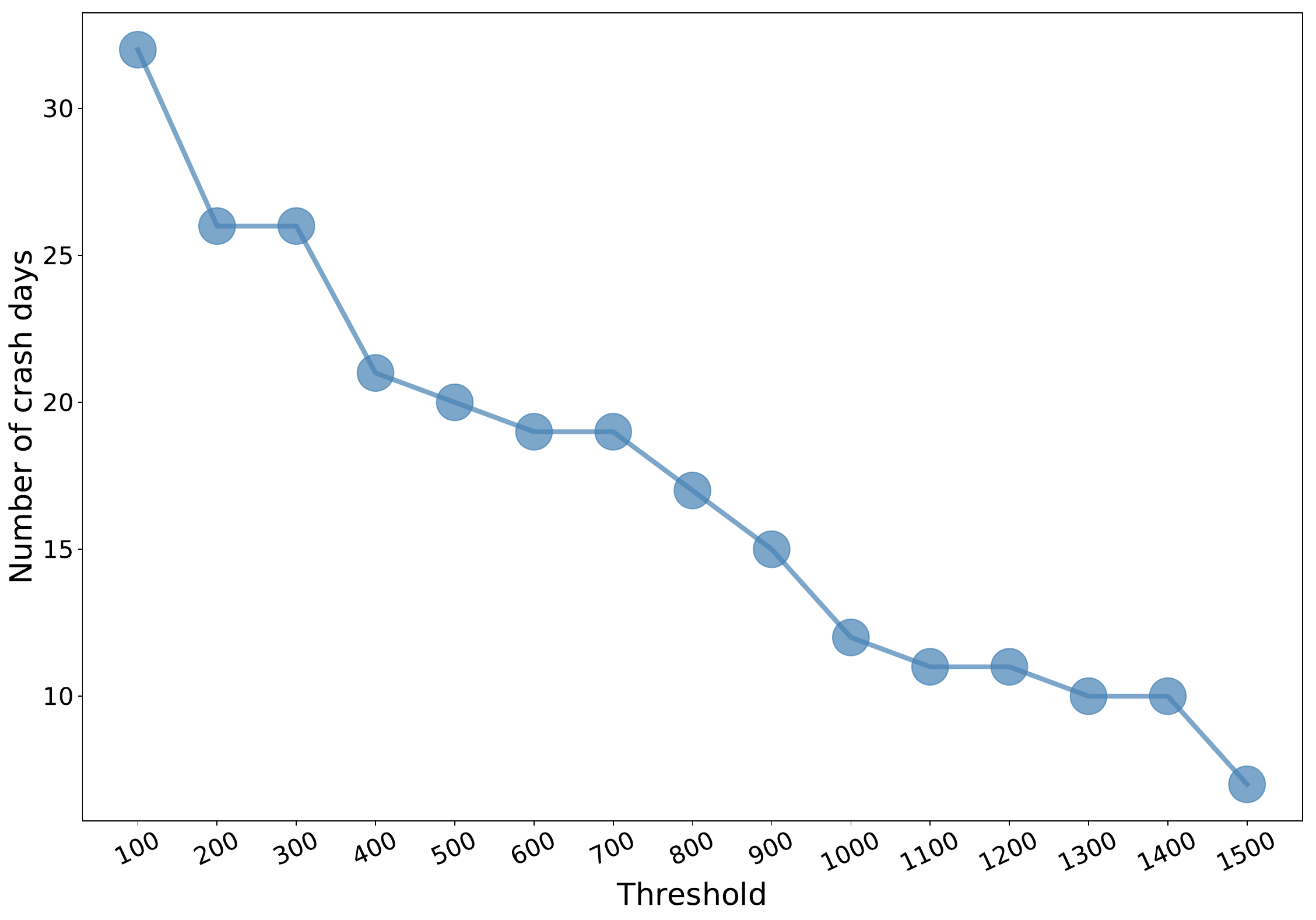} 
 	\caption{Threshold for the number of crash days. The abscissa indicates the threshold for identifying a market crash; if the number of stocks down to the limit is greater than or equal to this value on the day, it is defined as a market crash. }
  	\label{number_of_crash_days}
	\end{figure}
	
	\begin{figure}[htp]
			\centering
			\includegraphics[width=120mm]{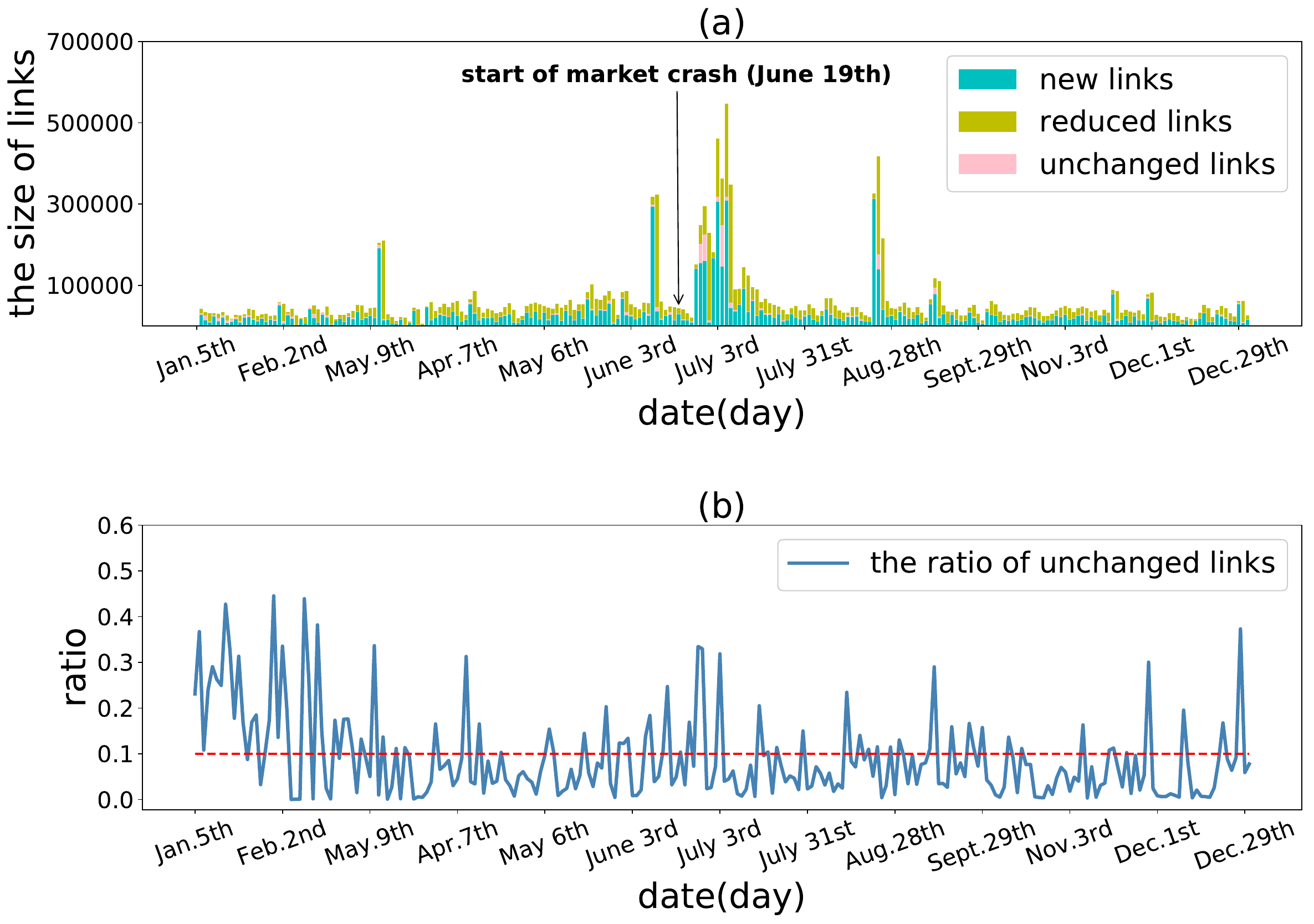}
			\caption{The evolution of links in illiquidity networks. (a) shows the size of new links, reduced links and unchanged links for two consecutive trading days. China's stock market evolves at a high frequency, especially on crash days. (b) shows the ratio of unchanged links, which indicates that only 10\% links remained on average for two consecutive trading days.}
			\label{fig:link} 
	\end{figure}
	
	\begin{figure}[!htp]
	\centering
	\includegraphics[width=0.9\textwidth]{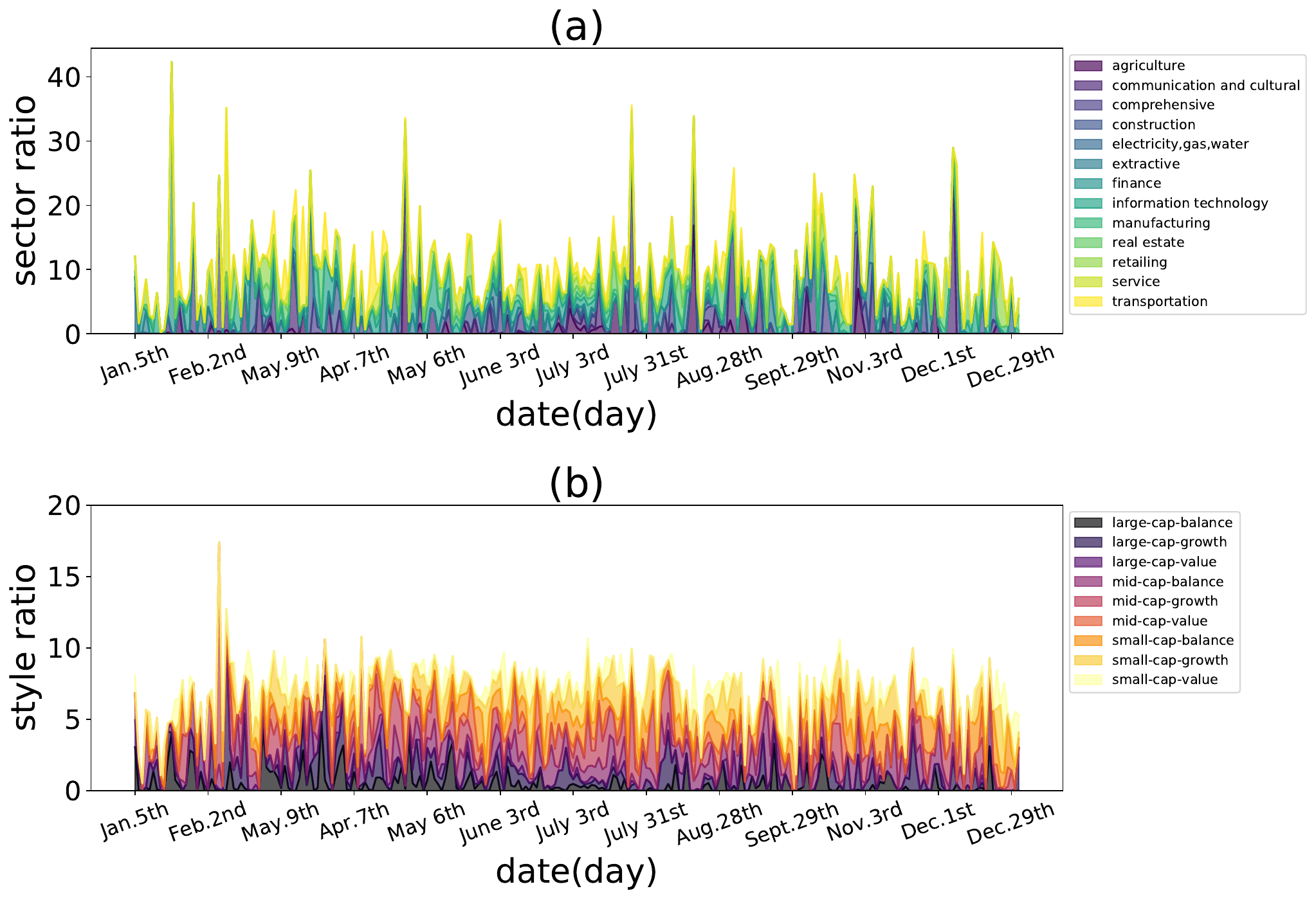} 
	\caption{ The betweenness centrality-weighted occurrence proportions of different sectors and capital styles in illiquidity networks. (a) shows the proportions of different sectors. (b) shows the proportions of different capital styles of stock values. }
	\label{betweenness_ratio}
	\end{figure}

	\begin{figure}[htp]
	\centering
	\includegraphics[width=0.9\textwidth]{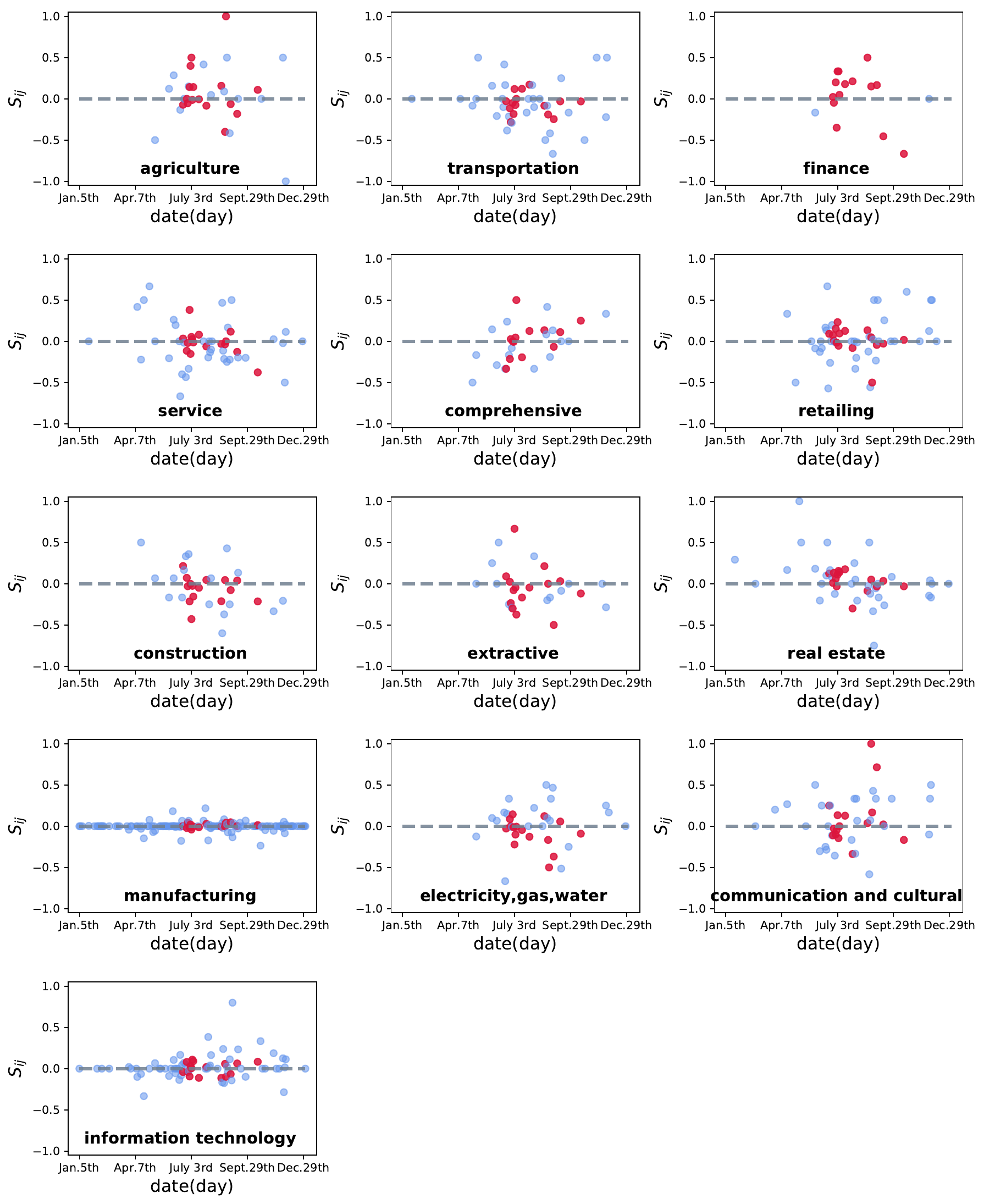} 
	\caption{ The significance of failing before peak using the method of betweenness centrality. The red dots indicate crash days, and the blue dots indicate noncrash days. As illustrated in the main text, $S_{ij}=R_{ji}^{bp}- R_{ji}^{bpr}$ and $S_{ij}$ may be positive or negative. If $S_{ij}$ is positive, then stocks within sector $i$ tend to fail before peaks. }
	\label{betweenness_sector_ratio}
	\end{figure}

	\begin{figure}[htp]
			\centering
			\includegraphics[width=120mm]{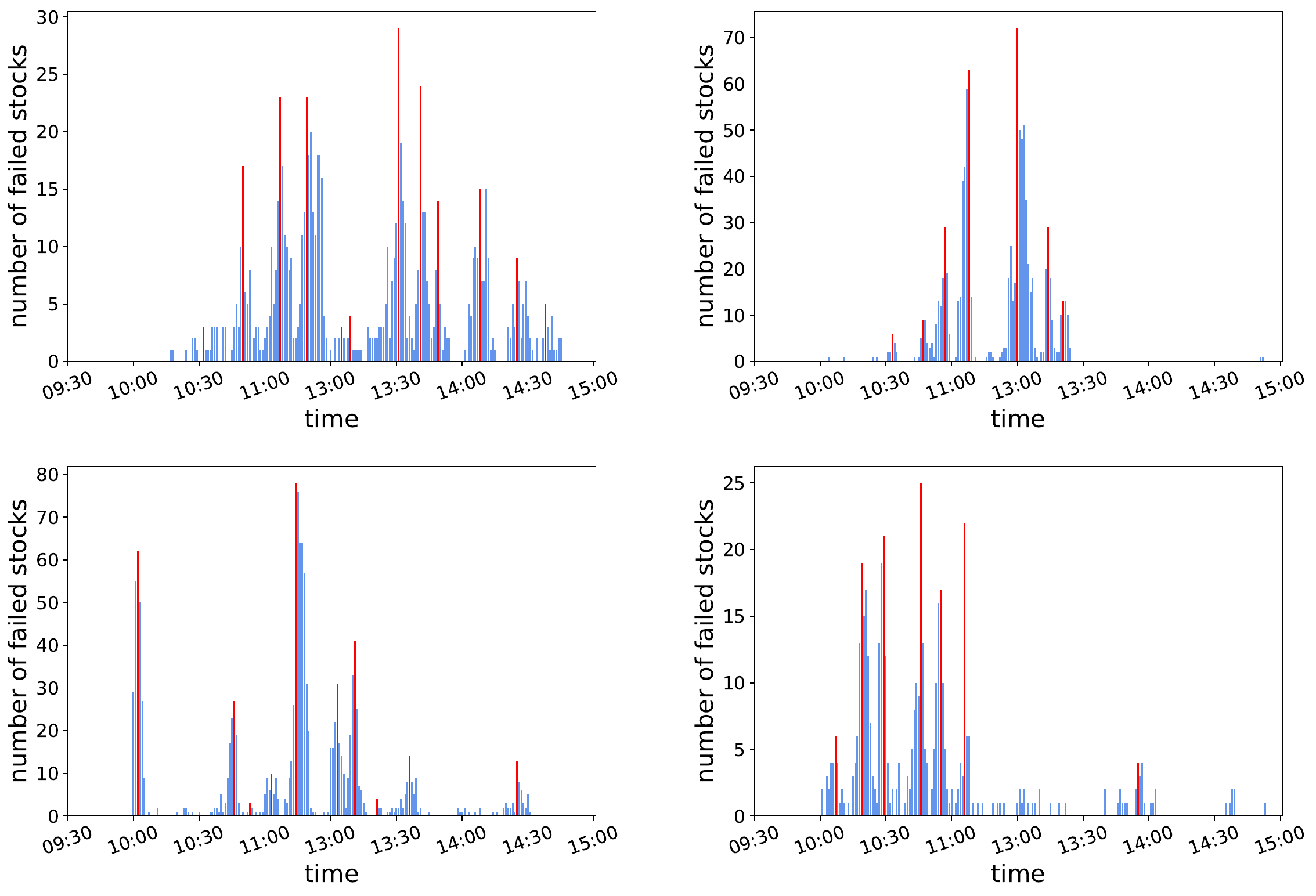}
			\caption{The peaks of stocks down to the limit. The principle of determining the peak is that the number of stocks mentioned above is the largest relative to the previous period and the subsequent period. There may be multiple peaks within a day. Note that, here, four crash days are selected to demonstrate the targeted peaks, including June 26th, June 29th, July 6th and July 7th, 2015.}
			\label{fig:peak2} 
	\end{figure}

	\begin{figure}[htp]
	\centering
	\includegraphics[width=1.0\textwidth]{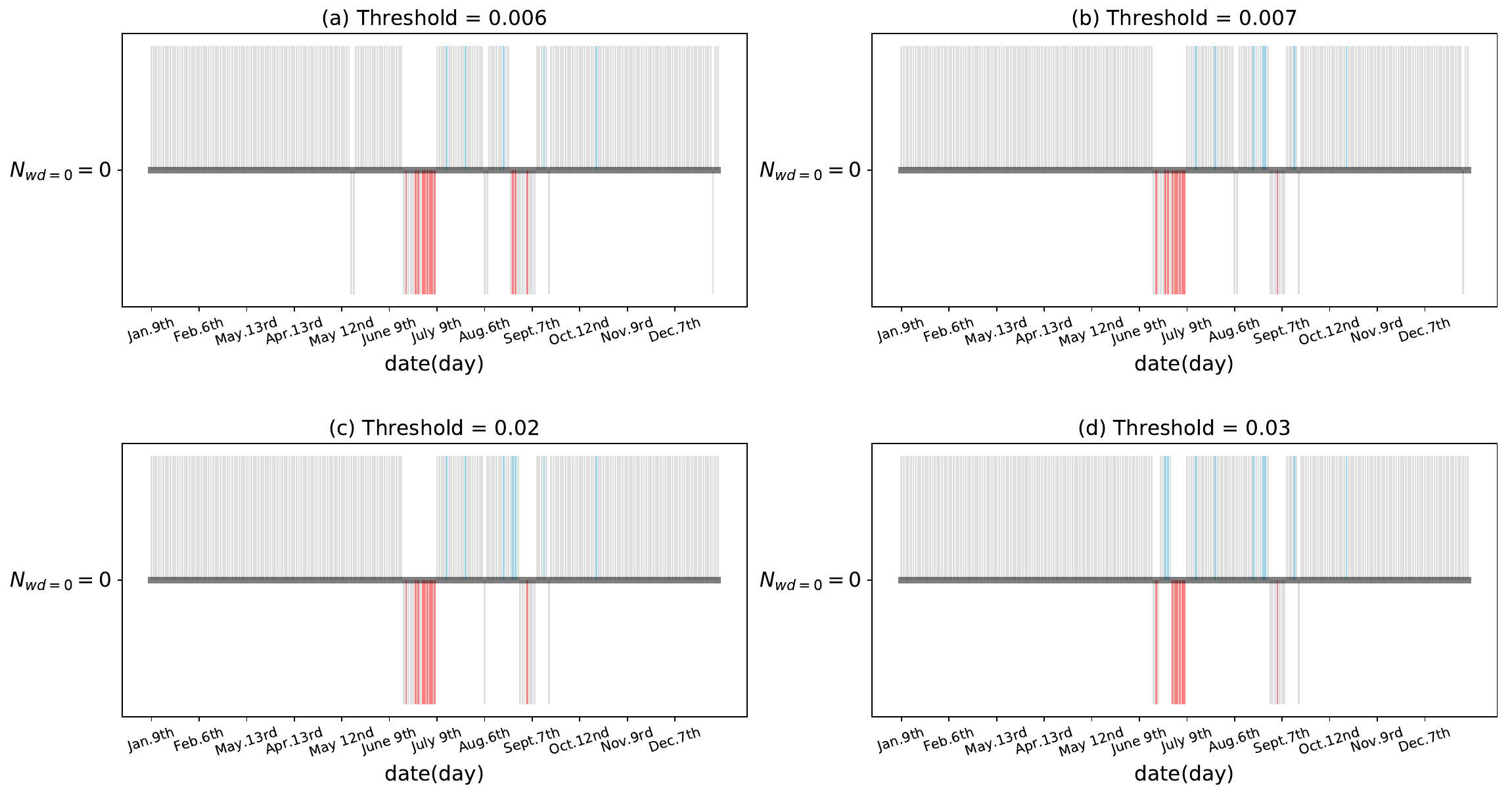} 
	\caption{The warning signal in illiquidity networks from different link thresholds. Similarly, gray bars represent noncrash days, red bars mark crash days that can be warned of one day in advance, and those that cannot be warned of in advance are colored blue. }
	\label{border_zero_bar}
\end{figure}

\begin{figure}[!h]
  \centering
  \includegraphics[width=0.9\textwidth]{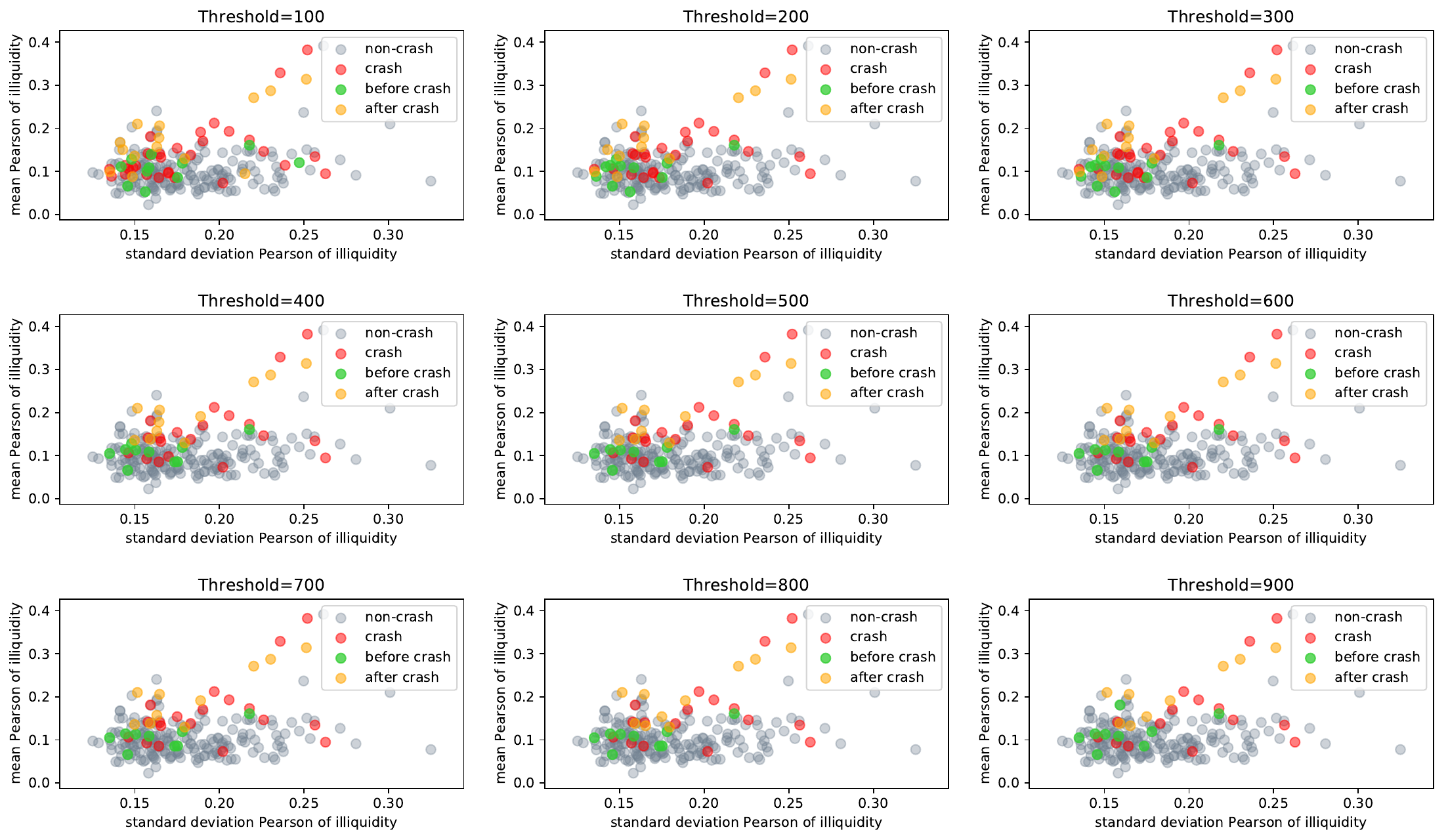} 
  \caption{ The mean and standard deviation of Pearson correlation-based networks. The threshold value is the criterion to determine market crash days, i.e., the number of stocks that go down to limit is greater than or equal to this value on this day. To demonstrate the robustness of our test, here, the results from several threshold values including 800 are separately shown. }
  \label{pearson_mean_std}
\end{figure}

\begin{figure}[htp]
	\centering
	\includegraphics[width=1.\textwidth]{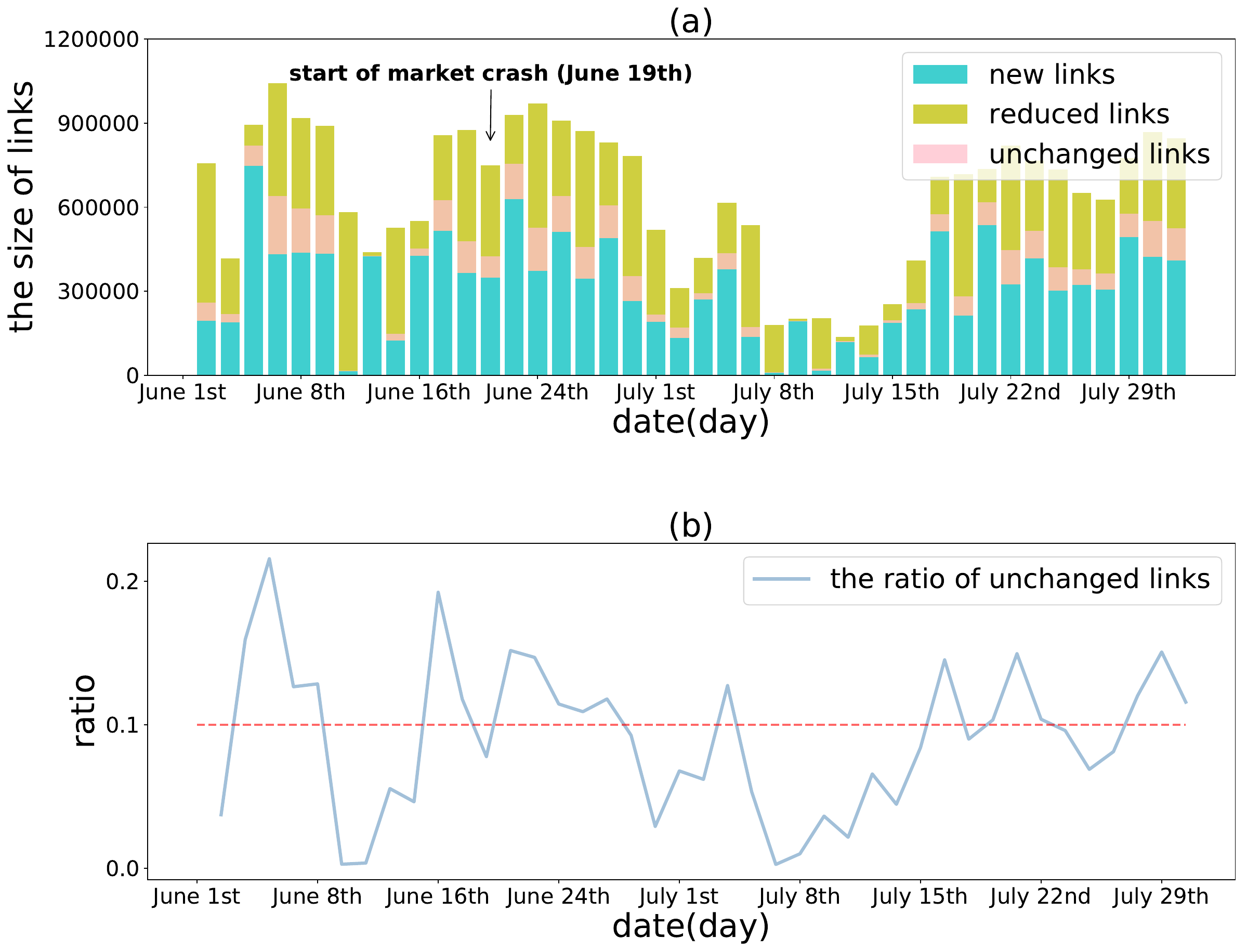} 
	\caption{ The evolution of links in Granger networks (time-lag=1, risk level=0.05). (a) shows the size of new links, reduced links and unchanged links for two consecutive trading days. Similar to illiquidity networks, China's stock market evolves at a high frequency even from the perspective of Granger networks, especially on crash days. (b) shows the ratio of unchanged links, which indicates that, similarly, only 10\% of links remained on average for two consecutive trading days. }
	\label{granger_edges}
\end{figure}

\begin{figure}[!h]
	\centering
	\includegraphics[width=1.\textwidth]{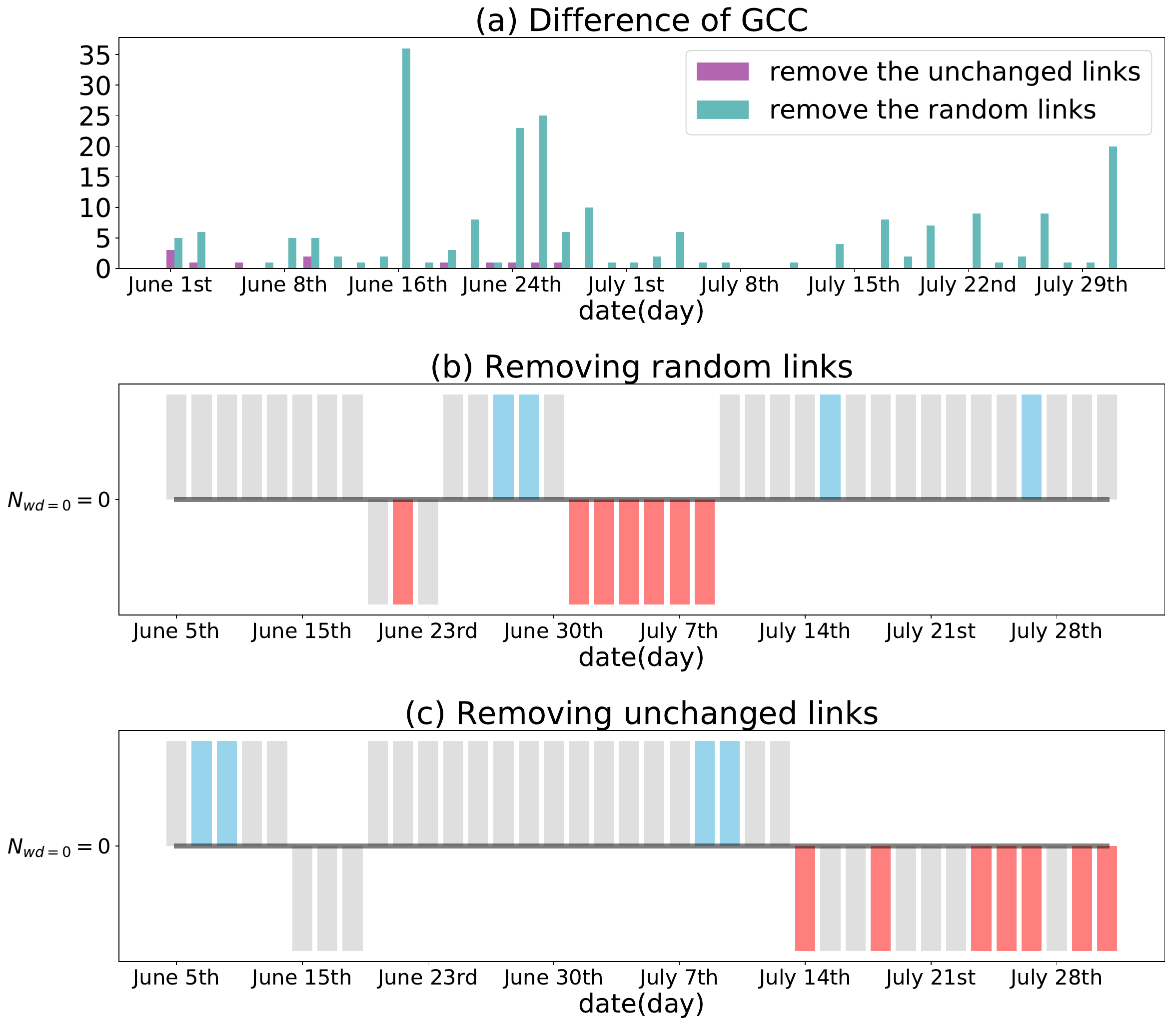} 
	\caption{The results after removing 10\% of links. (a) shows the difference between the GCC size of the initial network and the GCC of the network after removing the links. There are two ways to remove links: one is by removing 10\% of links, the other is removing links randomly, and the number of links is the same. (b) shows the early warning signal after removing the links randomly. (c) shows the early warning signal after removing the unchanged links. Removing these unchanged links does not break the network into pieces (GCC declines trivially, and the decline is even smaller than that from deleting links of the same volume at random), and the capability of detecting warning signals of crashes is sustained after the removal of these unchanged links. Note that, here, only the key months of June and July when market crashes occurred frequently are considered due to the heavy computational burden. }
	\label{removing_links}
\end{figure}

\end{appendices}

\end{document}